\documentclass{revtex4}

\usepackage{float}
\usepackage{graphicx}
\usepackage{latexsym}
\begin{document}
 \newcommand{\bq}{\begin{equation}}
 \newcommand{\eq}{\end{equation}}
 \newcommand{\bqn}{\begin{eqnarray}}
 \newcommand{\eqn}{\end{eqnarray}}
 \newcommand{\bqs}{\begin{equation}\begin{split}}
 \newcommand{\eqs}{\end{split}\end{equation}}
 \newcommand{\nb}{\nonumber}
 \newcommand{\lb}{\label}
 \newcommand{\p}{\partial}
 \newcommand{\tg}{\tilde{g}}
 \newcommand{\tc}{\tilde{\Gamma}}
\newcommand{\tdr}{\tilde{R}}
\newcommand{\rr}{\rangle}
\newcommand{\lr}{\langle}

\title{Post-selected Quantum Circuits}
\author{Michael Devin}

\begin{abstract}
The purpose of this paper is to show the unusual behavior of a number of simple circuits under the effects of post-selection. A useful duality exists between post-selected ensembles and a consistent picture of acausal physics embodying the application of Novikov's consistency postulate to the wave-function of a time machine. A competing view applies this postulate to density matrices instead, but that lies outside the scope of this paper. To find out the result of a measurement on a particular circuit, we consider a weighted distribution of histories between some initial and final time, such that each individually obeys ordinary quantum mechanics. The weight of a particular history is given by a joint distribution over the values of a part of the system dubbed the time machine, at two chosen times. In the dual post-selection picture the weight is simply the probability for the 'periodic bit' to satisfy the constraint, plus a small noise part, often assumed constant, representing the error rate of the time machine's channel. The emerging bit is one of a Bell pair, its partner kept in reserve to test that the bit entering the time machine later matches the one which emerged. This is done by interfering the incoming bit with the reserve member of the pair and keeping the experimental runs that interfere constructively. Noise can be effectively simulated by keeping a small proportion of runs that interfere destructively, or by perturbing the reserve bit to decohere it by a similar amount. Paradoxes remain but are tractable. 
\end{abstract}

\maketitle

\clearpage

The paper is organized into sections each focused on a particular set of circuits, and calculates the behavior of each according to several related models. The fully coherent or Bell state model, the decoherent model, and the noisy version of each. Each model can be described in terms of a correlation function, which provides the prescription for the relative weights of the sub-ensembles used in post-selection to simulate state projection dynamics. The example calculations are chosen to illustrate the unusual properties of post-selected systems, as well as familiarize the reader with the use of each model to calculate those behaviors. The Bell state model is perhaps the most well known\cite{seth1}. Also, a review of work on quantum teleportation may be of great help to readers otherwise unfamiliar with it\cite{tele1,vaid}.

\tableofcontents
\clearpage

\section{General formalism}

One approach to modeling time machines is through post-selected ensembles. 
The statistics of these ensembles can be equivalent to a deformation of quantum mechanics that includes projection and renormalization.
Treating time loops by selecting out inconsistent histories and renormalizing the wave-function can also be formulated this way.
The Bell state projection method consists of appending a maximally entangled pair of qubits, 
then projecting against the same pair state later. One bit is the reference bit, 
and the other the periodic channel or time machine state. The Bell state is given by
\bq
|\phi_{Bell}\rr = \frac{1}{\sqrt{2}}\left( |00\rr + |11\rr \right)
\eq
The out state that emerges from the loop is formed by the product of this with the environment.
\bq
|\psi_{B-out}\rr = |\phi_{Bell}\rr \otimes |\psi_{ex}\rr\\
\eq
The input state to the loop is obtained by normal unitary action of the circuit.
\bq
|\psi_{B-in}\rr = U_t |\phi_{B-out}\rr
\eq
Finally the state is projected against the same Bell state added in the first step to produce a reduced state vector.
\bq
|\bar{\psi}_B\rr = \lr \phi_{Bell} | \psi_{B-in}\rr
\eq
The norm of this projected state may be different from unity and so requires a renormalization factor to restore it.
\bqn
N^2 = \lr \bar{\psi}_B | \bar{\psi}_B \rr\\
|\psi_{final}\rr = N^{-1}|\bar{\psi}_B\rr
\eqn
As long as $N$ does not vanish, the evolution is unique and paradoxes are avoided. The value of $N$ however can have a physical meaning. It gives the relative frequency for the post-selection criteria to be met, and has measurable effects on the statistics of entanglement within the projected model.
This model has the advantage of mapping pure states to pure states, 
but fails for evolutions that map the Bell state bits onto one of the three orthogonal states.
\bqn
|\perp_1\rr =  \frac{1}{\sqrt{2}}\left( |00\rr - |11\rr \right)\\
|\perp_2\rr =  \frac{1}{\sqrt{2}}\left( |01\rr + |10\rr \right)\\
|\perp_3\rr  =  \frac{1}{\sqrt{2}}\left( |01\rr - |10\rr \right)
\eqn
These are produced by the action of a phase flip gate, a not gate and the combination of the two respectively.
 In each case $N$ vanishes.
Due to the condition that the external evolution $U_t$ leaves the reference bit unchanged, 
we can express the projected state as a sum of two other projected states. 
\bq
|\bar{\psi}_B\rr = |\bar{\psi_{00}}\rr + |\bar{\psi_{11}}\rr =  \lr 0 | U_t | 0,\psi_{ex} \rr + \lr 1 | U_t |1,\psi_{ex}\rr
\eq
The two components are each formed by appending only single channel eigenstate to $\psi_{ex}$ and then projecting against it again later.
The classical limit is formed by instead taking the decoherent mixture of these two component states. 
\bqn
\rho_{cl} = Z^{-1} \left( \bar{|\psi}_{00}\rr \lr \bar{\psi}_{00} | + |\bar{\psi}_{11}\rr\lr \bar{\psi}_{11} | \right)\\
Z = \lr \bar{\psi}_{00} | \bar{\psi}_{00} \rr +  \lr \bar{\psi}_{11} | \bar{\psi}_{11} \rr
\eqn
The classical partition function only vanishes if both terms vanish. The noisy periodic classical channel
can be described with a single bit error rate parameter $k$.
\bqn
|\bar{\psi}_{01}\rr = \lr 0 | U_t | 1,\psi_{ex}\rr \\
|\bar{\psi}_{10}\rr = \lr 1 | U_t | 0,\psi_{ex}\rr \\
Z = (1-k)\lr \bar{\psi}_{00} | \bar{\psi}_{00} \rr +  (1-k)\lr \bar{\psi}_{11} | \bar{\psi}_{11} \rr  + k \lr \bar{\psi}_{01} | \bar{\psi}_{01} \rr + k \lr \bar{\psi}_{10} | \bar{\psi}_{10} \rr \\
\rho_k = Z^{-1} \left( (1-k)| \bar{\psi}_{00} \rr\lr \bar{\psi}_{00} | +  (1-k)| \bar{\psi}_{11} \rr\lr \bar{\psi}_{11} |  + k | \bar{\psi}_{01} \rr\lr \bar{\psi}_{01} | + k | \bar{\psi}_{10} \rr\lr \bar{\psi}_{10} | \right)
\eqn
The most general approach is to consider a joint distribution over two states
\bqn
|\bar{\psi}(\phi_a,\phi_b) \rr = \lr \phi_a | U_t | \phi_b , \psi_{ex} \rr\\
Z = \int \omega( \phi_a,\phi_b) \lr \bar{\psi}(\phi_a,\phi_b)  |\bar{\psi}(\phi_a,\phi_b) \rr d\phi_a d\phi_b\\
\rho_\omega = Z^{-1}\int  \omega( \phi_a,\phi_b) |\bar{\psi}(\phi_a,\phi_b) \rr \lr\bar{\psi}(\phi_a,\phi_b) | d\phi_a d\phi_b
\eqn
Each combination of states $(\phi_a,\phi_b)$ represents a different history in the time machine picture, where a state $\phi_b$ emerges from the time machine channel and a state $\phi_a$ enters it after interaction with the environment via $U_t$. In the post-selection picture, we select histories to be a part of the ensemble based on a measurement of $\phi_a$ and $\phi_b$ at the appropriate times.
The above expression for the partition function $Z$ can be reduced to the sum over pairs of the eigenstates $e_i$.
\bq
Z = \sum_{ij} \omega_{ij} \lr e_i,\psi_{ex}| U_t^\dagger | e_j \rr\lr e_j | U_t | e_i \psi_{ex} \rr
\eq

The standard Bell model can be seen as a delta function choice for $\omega$
\bq
\omega_{Bell} = \delta(\phi_a - \phi_{Bell})\delta(\phi_b -\phi_{Bell})
\eq
And the noisy classical limit can be expressed as
\bqn
\omega_{Classical} = (1-k)\delta(\phi_a - 0)\delta(\phi_a - 0)+(1-k)\delta(\phi_a - 1)\delta(\phi_a - 1)+ \nb\\
k\delta(\phi_a - 0)\delta(\phi_a - 1)+k\delta(\phi_a - 1)\delta(\phi_a - 0)
\eqn
This description has a high degree of redundancy.
The full distribution function can be replaced by its moments multiplied by delta functions at each combination of eigenvalues.
If all of the moments are positive then the partition function should be non-vanishing for any unitary evolution $U_t$. To add noise to the Bell model we can add three delta functions at the three states orthogonal to the Bell state. The final density matrix of the system will be a mixture of four orthogonal projections.
\bqn
|\bar{\psi}_B\rr = \lr 00 + 11 | U_t | \phi_{Bell}, \psi_{ex} \rr\\
|\bar{\psi}_-\rr = \lr 00 - 11  | U_t | \phi_{Bell}, \psi_{ex} \rr\\
|\bar{\psi}_N\rr = \lr 01 + 10 | U_t | \phi_{Bell}, \psi_{ex} \rr\\
|\bar{\psi}_{N-}\rr \lr 01 - 10  | U_t | \phi_{Bell}, \psi_{ex} \rr\\
\lr \bar{\psi}_B|\bar{\psi}_B\rr + \lr \bar{\psi}_-|\bar{\psi}_-\rr + \lr \bar{\psi}_N|\bar{\psi}_N\rr + \lr \bar{\psi}_{N-}|\bar{\psi}_{N-}\rr =1
\eqn
To simulate the standard depolarizing channel we use  a partition function of
\bq
Z_{\lambda} = (1-\lambda) \lr\bar{\psi}_B |\bar{\psi}_B\rr + \lambda/4
\eq
and a density matrix of
\bqn
\rho_\lambda = Z^{-1}(1-\frac{3\lambda}{4}) |\bar{\psi}_B \rr\lr \bar{\psi}_B | + \frac{\lambda}{4Z}  |\bar{\psi}_- \rr\lr \bar{\psi}_- |\nb\\
+ \frac{\lambda}{4Z}  |\bar{\psi}_N \rr\lr \bar{\psi}_N |+ \frac{\lambda}{4Z}  |\bar{\psi}_{N-} \rr\lr \bar{\psi}_{N-} |
\eqn
The classical noisy channel can be obtained by the choice
\bqn
Z_{cl} = (1-k)\lr \bar{\psi}_B|\bar{\psi}_B\rr + (1-k)\lr \bar{\psi}_-|\bar{\psi}_-\rr + k\lr \bar{\psi}_N|\bar{\psi}_N\rr + k\lr \bar{\psi}_{N-}|\bar{\psi}_{N-}\rr\\
\rho_{cl} = \frac{1-k}{Z} |\bar{\psi}_B \rr\lr \bar{\psi}_B | +\frac{1-k}{Z}  |\bar{\psi}_- \rr\lr \bar{\psi}_- |\nb\\
+ \frac{k}{Z}|\bar{\psi}_N \rr\lr \bar{\psi}_N |+ \frac{k}{Z} |\bar{\psi}_{N-} \rr\lr \bar{\psi}_{N-} |
\eqn
Where the equal weight given to $\bar{\psi}_-$ decoheres the state along the $0-1$ basis.
If we choose the weight function to be a constant then ensemble will be unskewed and the resulting statistics those of a completely random channel.
\bqn
Z_{flat} = \int \lr \phi_b \psi_{ex} | U_t^\dagger | \phi_a \rr\lr \phi_a | U_t | \phi_b \psi_{ex} \rr d\phi_a d\phi_b\nb\\
 =\sum_{ij}^d |\lr e_j | U_t | e_i\psi_{ex}\rr|^2\left((2\pi)^{d-1}\int_0^\pi\int_{S_{d-2}} \cos^2\theta \sin^{d-2}\theta d\theta \cdot dS_{d-2}\right)^2 \nb\\
= (2\pi)^{2d-2}S_{d-1}^2 \frac{1}{d}\\
\rho_{flat} = Z_{flat}^{-1}\int \lr \phi_a | U_t | \phi_b \psi_{ex} \rr\lr \phi_b \psi_{ex} | U_t^\dagger | \phi_a \rr d\phi_a d\phi_b \nb\\
= Z^{-1} \sum_{ij}^d \lr e_j|U_t|e_i\psi_{ex}\rr\lr e_i \psi_{ex}|U_t^\dagger|e_j\rr\left((2\pi)^{d-1}\int_0^\pi\int_{S_{d-2}} \cos^2\theta \sin^{d-2}\theta d\theta \cdot dS_{d-2}\right)^2 \nb\\
= \frac{1}{d}\sum_{ij}^d \lr e_j|U_t|e_i\psi_{ex}\rr\lr e_i \psi_{ex}|U_t^\dagger|e_j\rr
\eqn
for the $n$-qubit channel. Other notable choices of weight function are
\bq
\omega_{quad}(\phi_a,\phi_b) = |\lr\phi_a|\phi_b\rr|^2\\
\eq
and
\bq
\omega_\delta(\phi_a,\phi_b) = \delta(\phi_a-\phi_b)
\eq
This gives mixtures defined by
\bqn
Z_{quad}= \int |\lr \phi_a|\phi_b\rr|^2 \lr \phi_b \psi_{ex} | U_t^\dagger | \phi_a \rr\lr \phi_a | U_t | \phi_b \psi_{ex} \rr d\phi_a d\phi_b \nb\\
=\sum_{ij}^d \omega_{ij} \lr e_i \psi_{ex} | U_t^\dagger | e_j \rr\lr e_j | U_t | e_i \psi_{ex} \rr\\
\omega_{ij}=\frac{2\delta_{ij} +1}{d(d+2)}(2\pi)^{2d-2}S_{d-1}^2\\
\rho_{quad}= Z^{-1}\sum_{ij}^d \omega_{ij} \lr e_j | U_t | e_i \psi_{ex} \rr \lr e_i \psi_{ex} | U_t^\dagger | e_j \rr
\eqn
for the inner product correlated channel and 
\bqn
Z_\delta = \int \delta(\phi_a-\phi_b) \lr \phi_b \psi_{ex} | U_t^\dagger | \phi_a \rr\lr \phi_a | U_t | \phi_b \psi_{ex} \rr d\phi_a d\phi_b \nb\\
=\sum_{ij}^d \omega_{ij} \lr e_i \psi_{ex} | U_t^\dagger | e_j \rr\lr e_j | U_t | e_i \psi_{ex} \rr\\
\omega_{ij} = \frac{1}{d}\delta_{ij}(2\pi)^{d-1}S_{d-1}\\
\rho_\delta = Z^{-1} \sum_i^d \omega_{ij}\lr e_i | U_t | e_i\psi_{ex}\rr\lr e_i\psi_{ex}|U_t^\dagger | e_i \rr
\eqn
for the delta correlated channel. In the above expressions $S_d$ refers to the area of the unit $d-$sphere. The powers of $2\pi$ come from the mutual phases between components of the $d-$dimensional complex unit vectors.  The spread of the correlation function then gives us the effective noise of the channel.
Later we will see that, $\omega_\delta$ gives us a channel with noise of $1/2$. 
We will be using the noisy Bell and noisy classical channels for the majority of this paper.

\section{Simple loop}

This section covers the action of the bare channel. The circuit is a simple swap operation between the periodic channel and an external channel.
The treatment in the Bell state model begins with an arbitrary input qubit,
\bq
|\psi_1\rr = \alpha|0\rr + \beta|1\rr
\eq
 we can add a maximally entangled pair in a product state with it.
\bq
|\phi_{Bell}\rr = \frac{1}{\sqrt{2}}( |00\rr + e^{i\theta}|11\rr)
\eq
with the left qubit as our reference bit, and the right qubit as the time machine 'out' state.
The combined state before acting with the rest of the circuit is simply
\bq
|\psi_{out}\rr = |\phi_{Bell}\rr \otimes |\psi_1\rr = \frac{1}{\sqrt{2}}(\alpha|000\rr + \alpha e^{i\theta}|110\rr +\beta|001\rr +\beta e^{i\theta}|111\rr)
\eq
Now to simulate sending channel $\psi_1$ into the time machine and observing what comes out, we swap the in/out channel with channel 1.
\bq
U_{swap} = \psi_1 \leftrightarrow \phi
\eq
The state after the swap is now,
\bq
|\psi_{in}\rr = U_{swap}|\psi\rr  =  \frac{1}{\sqrt{2}}(\alpha|000\rr + \alpha e^{i\theta}|101\rr +\beta|010\rr +\beta e^{i\theta}|111\rr)
\eq
Generally the reference bit will be unchanged by the circuit, unless we wish to model some internal dynamics of the time machine.
Now we project against the original product state $\phi_{Bell}$ removing the first two bits and leaving only the channel $\psi_1$.
\bq
|\bar{\psi}\rr =  \frac{1}{2}(\lr 00,x|\psi'\rr + e^{-i\theta}\lr 11,x |\psi'\rr) = \frac{1}{2}(\alpha|0\rr + \beta|1\rr) = \frac{1}{2}|\psi_1\rr = N|\psi_1\rr 
\eq 
The $,x$ notation is a reminder that the third qubit is not contracted, but remains. 
This is none other than an ordinary quantum teleportation in the case that the final measurement against $\lr \phi_{Bell} |$ returned 1.
 The the square of the factor N represents the portion of ensembles that satisfy the selection condition represented by the state projection.\\

The phase of the Bell pair drops out of the final calculation but other representations are possible.
Consider a rotated pair as a reference state,
\bq
|\phi_{alt}\rr = \frac{1}{2}\left( |00\rr + |01\rr + |10\rr - |11\rr \right)
\eq
The calculation is more tedious, but gives the same result. 
\bqn
|\psi_{out}\rr = |\phi_{alt}\rr \otimes |\psi_1\rr = \frac{1}{2}( \alpha|000\rr + \alpha|010\rr + \alpha|100\rr - \alpha|110\rr +\beta|001\rr + \beta|011\rr + \beta|101\rr - \beta|111\rr )\\
|\psi_{in}\rr = U_{swap} |\psi\rr = \frac{1}{2}( \alpha|000\rr + \alpha|001\rr + \alpha|100\rr - \alpha|101\rr +\beta|010\rr + \beta|011\rr + \beta|110\rr - \beta|111\rr )\\
|\bar{\psi}\rr =  \frac{1}{2}\left( \lr 00,x| + \lr 01,x| + \lr 10,x| - \lr 11,x| \right) |\psi'\rr = \frac{1}{2} (\alpha |0\rr + \beta|1\rr)\\
N=1/2
\eqn
By using a non-maximally entangled pair we get a distorted signal. Consider the case
\bq
|\phi_{twist}\rr = \frac{1}{\sqrt{2}}|00\rr + \frac{1}{2}|01\rr + \frac{1}{2}|11\rr
\eq
The product state after channel swapping is
\bq
|\psi_{in}\rr = \frac{\alpha}{\sqrt{2}}|000\rr + \frac{\alpha}{2}|001\rr + \frac{\alpha}{2}|101\rr + \frac{\beta}{\sqrt{2}}|010\rr + \frac{\beta}{2}|011\rr + \frac{\beta}{2}|111\rr
\eq
The projection of this state against $\lr \phi_{twist} |$ gives
\bq
|\bar{\psi}\rr = \frac{1}{2}|\psi_1\rr + \frac{\beta}{\sqrt{8}}|0\rr + \frac{\alpha}{\sqrt{8}}|1\rr
\eq
So it is important in this case for the pair to be maximally entangled.
We can also consider a decoherent time machine model. In this case the out state is again an arbitrary product, but is a single qubit rather than a pair.
\bq
|\phi_{TM}\rr = \gamma_0 |0\rr + \gamma_1 |1\rr = \cos{\theta}|0\rr + e^{i\xi}\sin{\theta}|1\rr
\eq
Taking the product and then the swap operation, we have
\bqn
|\psi_{out}\rr = |\phi_{TM}\rr \otimes |\psi_1\rr = \alpha\gamma_0|00\rr + \alpha\gamma_1|10\rr + \beta\gamma_0|01\rr + \beta\gamma_1|11\rr\\
|\psi_{in}\rr = U_{swap} |\psi\rr =  \alpha\gamma_0|00\rr + \alpha\gamma_1|01\rr + \beta\gamma_0|10\rr + \beta\gamma_1|11\rr\\
|\bar{\psi}\rr = (\gamma_0^{\dagger}\lr 0,x| + \gamma_1^{\dagger}\lr 1,x| ) |\psi'\rr\\= (\alpha\gamma_0^{\dagger}+\beta\gamma_1^{\dagger})( \gamma_0|0\rr + \gamma_1|1\rr)\nb\\
\lr \phi_{TM}| \psi_1 \rr \cdot |\phi_{TM}\rr
\eqn
Summing over an orthogonal set of states for $\phi_{TM}$ in this single qubit method is identical to the Bell state method mentioned earlier, if we fix the mutual phase of the terms. Here, the final state is the time machine out state, with a weight corresponding to how well channel $\psi_1$ matches the in state $\phi_{TM}$. Checking this with a coherent integration of $\bar{\psi}$ over $\phi_{TM}$ we have,
\bqn
|\bar{\psi}\rr = \bar{\alpha}|0\rr + \bar{\beta}|1\rr\\
\bar{\alpha} = \alpha \cos^2{\theta} + \beta e^{-i\xi}\sin{\theta} \cos{ \theta} \\
\bar{\beta} = \alpha e^{i\xi}\sin\theta\cos\theta + \beta\sin^2\theta\\ 
Z = \int d\phi = 2\pi^2\\
\int \bar{\alpha}  d\phi = \pi^2 \alpha\\
\int \bar{\beta}  d\phi = \pi^2 \beta \\
Z^{-1}\int |\bar{\psi}\rr d\phi = \frac{1}{2}( \alpha|0\rr + \beta|1\rr)= \frac{1}{2}|\psi_1\rr
\eqn
This should not be surprising since the operations here are all linear, and the contributions from rotated states will simply be linear combinations of the contributions from the basis states, providing the overall constant factor $Z$.\\ We can also form decoherent mixtures of the projected states $\bar{\psi}$.
\bqn
\rho = Z^{-1}\int |\bar{\psi}\rr\lr \bar{\psi} |  d\phi_{TM}\\
 Z= \int | \lr \phi_{TM},x | \psi_{in}\rr |^2 d\phi_{TM} =\int \lr \bar{\psi}|\bar{\psi}\rr d\phi =  \int  N_\phi N_\phi^\dagger d\phi
\eqn
This prescription comes from the $\omega_\delta$ correlation function introduced earlier. The system in this model is an ensemble average of decoherent histories. Each of those histories is characterized by what state emerges from the time machine, and weighted by the norm squared of the projection of the full state against the emitted state.
In the simple loop, this choice of weight function integrates to
\bq
Z = \int_0^\pi \int_0^{2\pi}[ \alpha^2\cos^2\theta + \beta^2\sin^2\theta + 2\alpha\beta\sin\theta\cos\theta\cos\xi ]d\theta d\xi = \pi^2
\eq
 The density matrix terms are
\bqn
\rho_{00} = Z^{-1} \int [\omega(\theta,\xi)\cos^2\theta]d\theta d\xi = \frac{1}{2}\alpha^2+\frac{1}{4}\\
\rho_{11} = Z^{-1}\int [\omega(\theta,\xi)\sin^2\theta]d\theta d\xi = \frac{1}{2}\beta^2+\frac{1}{4}\\
\rho_{10}=\rho_{01}^\dagger = \frac{1}{4}\alpha\beta
\eqn
which can be expressed more simply as
\bq
\rho_{TM} = \frac{1}{2} |\psi_1\rr\lr \psi_1 | + \frac{1}{4}I
\eq
So this weight function gives a channel with a $1:1$ signal noise ratio. 
The $\delta$-weight model is also investigated by \cite{mark}, under the formalism called T-CTCs. 
The treatment of the unproven proof in the $\delta$-model also loses coherence, and can be thought of as a realization of the mixing guessed by Hawking in \cite{hawk}.
The two qubit simple loop circuit can be expressed in the Bell state model by taking the product of the in state with a maximally entangled pair of 2 qubit Hadamard states.
The projection state is given by
\bq
|\phi_{TM}\rr = \frac{1}{2}\left( |0000\rr + |0101\rr + |1010\rr + |1111\rr\right) 
\eq
Defining the input state to be transported as
\bq
|\psi_1\rr = \gamma_{00}|00\rr + \gamma_{01}|01\rr + \gamma_{10}|10\rr + \gamma_{11}|11\rr
\eq
Then the out state is simply
\bq
|\psi_{out}\rr = |\phi_{TM}\rr \otimes |\phi_1\rr
\eq
The swap operation gives us the in state that will be projected against $\phi_{TM}$.
\bq
|\psi_{in}\rr = U_{swap52}U_{swap64}|\psi_{out} \rr
\eq
Projecting gives the state
\bq
|\bar{\psi}\rr = \lr \phi_{TM} | \psi_{in}\rr = \frac{1}{4}|\psi_1\rr
\eq
The classical limit will decohere the incoming channel along the time machine's preferred basis. To see this we consider the weights. 
First the different out states will be defined by the product of $\psi_1$ with each eigenstate of the time machine.
\bq
|\psi_{out}(i)\rr = |e_i\rr \otimes |\psi_1\rr
\eq
and the in states in now acting only on 4 qubits so that the swap operation is
\bq
|\psi_{in}(i)\rr = U_{swap13}U_{swap24}|\psi_{out}(i)\rr 
\eq
Now the weight of each eigenstate will be
\bqn
\omega_{00} = k/4 + (1-k)|\lr 00,xx |\psi_{in}(00)\rr|^2 = k/4 +(1-k)\gamma_{00}\gamma_{00}^\dagger \\
\omega_{01} = k/4 + (1-k)|\lr 01,xx |\psi_{in}(01)\rr|^2 = k/4 +(1-k)\gamma_{01}\gamma_{01}^\dagger \\
\omega_{10} = k/4 + (1-k)|\lr 10,xx |\psi_{in}(10)\rr|^2 = k/4 +(1-k)\gamma_{10}\gamma_{10}^\dagger \\
\omega_{11} = k/4 + (1-k)|\lr 11,xx |\psi_{in}(11)\rr|^2 = k/4 +(1-k)\gamma_{11}\gamma_{11}^\dagger
\eqn
where we have renormalized the noise constant to reflect the larger number of eigenstates. This gives the classical partition function as
\bq
Z_{cl} = k + (1-k)(\gamma_{00}\gamma_{00}^\dagger+\gamma_{01}\gamma_{01}^\dagger+\gamma_{10}\gamma_{10}^\dagger+\gamma_{11}\gamma_{11}^\dagger) = 1
\eq
And an output density matrix of
\bqn
\rho_{cl} = \omega_{00}|00\rr\lr0 0| +\omega_{01}|01\rr\lr 01| +\omega_{10}|10\rr\lr 10| +\omega_{11}|11\rr\lr 11| \nb\\
= \frac{k}{4}I + (1-k)diag (\gamma_{00}\gamma_{00}^\dagger ,\gamma_{01}\gamma_{01}^\dagger ,\gamma_{10}\gamma_{10}^\dagger ,\gamma_{11}\gamma_{11}^\dagger )
\eqn

 With a few basics out of the way, we now proceed to examine the most iconic scenario in acausal physics, the well known grandfather paradox.


\section{Grandfather circuit}

This and the faulty gun circuit are also discussed in \cite{seth2}, and included here for comparison and commentary.
One of the most basic and important circuits to analyze is the quintessential causal paradox, where whichever state emerges, an orthogonal state is sent in by the action of $U_t$. Beginning with the Bell state
\bq
|\phi_{Bell}\rr = \frac{1}{\sqrt{2}}( |00\rr + |11\rr)
\eq
the paradox circuit is defined by acting on $\phi_{out}$ with a simple not gate. To begin with we will consider any external system to remain in a product state.
\bq
|\psi_{in}\rr = U_{not2}|\psi_{out}\rr = U_{not2}|\phi_{Bell}\rr \otimes |\psi_{ext}\rr= \frac{1}{\sqrt{2}}( |01\rr + |10\rr) \otimes |\psi_{ext}\rr
\eq
Where $U_{not2}$ implements a not operation on the second bit of the Bell pair. In general we will assume the first bit of the Bell pair to be unchanged by any circuit.
The projection operation now gives zero, indicating that the selection criteria are never met. 
\bq
\lr \phi_{TM} | \psi_{in}\rr = 0 \otimes |\psi_{ext}\rr
\eq
\begin{figure}[h!]
  \caption{Grandfather circuit with decohering measurements.}
  \centering
    \includegraphics[width=0.4\textwidth]{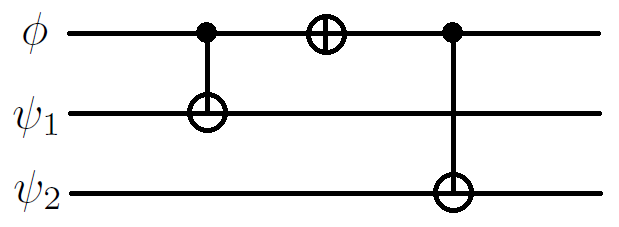}
\end{figure}

To better understand the situation, we can look at a perturbation of the state, either with noise or with a small rotation away from an exact not gate.
First consider the perturbation of $U_t$ such that
\bq
\tilde{U}_t=(1-\epsilon)U_{not} + \epsilon I
\eq
Now we have
\bq
|\tilde{\psi}\rr = \tilde{U}_t |\psi\rr = \left[\frac{1-\epsilon}{\sqrt{2}}\left(|01\rr+|10\rr \right) + \frac{\epsilon}{\sqrt{2}}\left( |00\rr + |11\rr \right)\right] \otimes |\psi_{ext}\rr
\eq 
Now the projection gives
\bq
\lr \phi_{TM} | \tilde{\psi} \rr = \epsilon |\psi_{ext}\rr
\eq
and so
\bq
N=\epsilon
\eq
We can now renormalize $\bar{\psi}$, but there is still a problem.
A unique limit does not exist in this case. Different perturbations of $U_{not}$ will give different projections, as the direction of perturbation will determine the direction of $\bar{\psi}$. Consider a different perturbation
\bq
\tilde{U}_t=(1-\epsilon)U_{not} + \epsilon U_\xi
\eq
this projection gives
\bq
\lr \phi_{TM} | \tilde{\psi} \rr = \epsilon U_\xi|\psi_{ext}\rr
\eq
Paradoxes are singular points in the behavior of the projection. The closer one gets to a circuit that has a vanishing normalization factor, the more sensitive the projection result becomes to small perturbations.
In general we will see that the magnitude of the normalization factor will be a measure of how far from normal the behavior of the system will be. The same result can be found with a noise term. Here we perturb the projection operator.
\bq
\lr \phi_{TM}|_{eff} = \frac{1-k}{\sqrt{2}}\left(\lr 00| + \lr 11| \right) + e^{i\vartheta}\frac{k}{\sqrt{2}}\left( \lr 01| + \lr 10|\right)
\eq
This represents a small amplitude for the qubit to flip during 'transit', such that
\bq
N=k
\eq
If we take the classical limit for a decoherent time machine, the partition function gives
\bq
Z = \epsilon + \int \omega(\phi) d\phi = \epsilon +  |\lr 0| U_{not} | 0\rr |^2 + | \lr 1 | U_{not} | 1\rr |^2 = \epsilon
\eq
This gives still gives zero for the density matrix unless the weight function is also perturbed, in which case 
\bq
\rho = Z^{-1} \left( \omega(|0\rr)|0\rr\lr 0| + \omega(|1\rr) |1\rr\lr 1| \right) = \frac{1}{2}I
\eq
For a partially decoherent weight model such as
\bq
\omega = | \lr \phi,x | U_{not} | \phi\otimes\psi \rr |^2
\eq
where $\phi$ is given by
\bq
|\phi\rr = \gamma_0 |0\rr + \gamma_1 e^{i\xi} | 1 \rr = \cos \theta |0\rr + \sin \theta e^{i\xi}|1\rr
\eq
now
\bq
\omega_{not}(\phi) = 4\cos^2 \theta \sin^2 \theta \cos^2 \xi
\eq
for a simple not gate or
\bq
\omega_{rot}(\phi) = 4\cos^2 \theta \sin^2 \theta \sin^2 \xi
\eq
for a not and phase flip. 
This gives us a partition function of
\bq
Z_{not}=Z_{rot} = \frac{\pi^2}{2}
\eq
and density matrix of
\bq
\rho_{not}=\rho_{rot} = Z^{-1}\int \omega |\bar{\psi}\rr\lr \bar{\psi} | = \frac{1}{2}(|0\rr\lr 0| + |1\rr\lr 1|)
\eq
in more detail for $\rho_{not}$,
\bqn
\rho_{00} = \frac{2}{\pi^2}\int_0^\pi \int_0^{2\pi} 4[\cos^4\theta\sin^2\theta\cos^2\xi ]d\theta d\xi =\frac{1}{2}\\
\rho_{01} =  \frac{2}{\pi^2}\int_0^\pi \int_0^{2\pi} 4e^{i\xi}[\cos^3\theta\sin^3\theta\cos^2\xi  ]d\theta d\xi =0\\
\rho_{11} = \frac{2}{\pi^2}\int_0^\pi \int_0^{2\pi} 4[\cos^2\theta\sin^4\theta\cos^2\xi ]d\theta d\xi =\frac{1}{2}\\
\rho_{10} =  \frac{2}{\pi^2}\int_0^\pi \int_0^{2\pi} 4e^{-i\xi}[\cos^3\theta\sin^3\theta\cos^2\xi  ]d\theta d\xi =0
\eqn
And for $\rho_{rot}$ ,
\bqn
\rho_{00} = \frac{2}{\pi^2}\int_0^\pi \int_0^{2\pi} 4[\cos^4\theta\sin^2\theta\sin^2\xi ]d\theta d\xi =\frac{1}{2}\\
\rho_{01} =  \frac{2}{\pi^2}\int_0^\pi \int_0^{2\pi} 4e^{i\xi}[\cos^3\theta\sin^3\theta\sin^2\xi  ]d\theta d\xi =0\\
\rho_{11} = \frac{2}{\pi^2}\int_0^\pi \int_0^{2\pi} 4[\cos^2\theta\sin^4\theta\sin^2\xi ]d\theta d\xi =\frac{1}{2}\\
\rho_{10} =  \frac{2}{\pi^2}\int_0^\pi \int_0^{2\pi} 4e^{-i\xi}[\cos^3\theta\sin^3\theta\sin^2\xi  ]d\theta d\xi =0
\eqn
In each decoherent model, the paradox maximizes entropy for the circulating qubit. Another implementation of the paradox is the controlled phase flip circuit. It is essentially another rotation of the previous case, included here for completeness.
\bq
U_{cpf} = |x,0\rr\lr x,0| - |x,1\rr\lr x,1|
\eq
The joint state in the standard Bell  representation is 
\bq
|\psi_{out}\rr = |\phi_{Bell}\rr\otimes |\psi_1\rr = \frac{1}{\sqrt{2}}( |00\rr + |11\rr ) \otimes ( \alpha |0\rr + \beta | 1\rr )
\eq
The phase flip gives the in state as
\bq
|\psi_{in}\rr = U_{cpf}|\psi_{out}\rr = \frac{\alpha}{\sqrt{2}} (|00\rr + |11\rr)\otimes |0\rr + \frac{\beta}{\sqrt{2}} ( |00\rr - |11\rr ) \otimes |1\rr
\eq 
The projection is then
\bq
\lr \phi_{Bell}^\dagger , x| \psi_{in}\rr = \alpha|0\rr
\eq
Destructive interference in the looping qubit removes the part of the state along $|1\rr$.
 If $\alpha$ vanishes then the grandfather paradox occurs.
A third implementation of the grandfather circuit is available by combining the phase flip and not gates. 
These three grandfather circuits represent mappings that take the initial Bell state to one of the three orthogonal entangled states.
 In the classical case, only the not circuit realizes the paradox, since the classical limit makes the weight of the post-selected ensemble indifferent to the mutual phase between the classical eigenstates.
Testing the behavior of the controlled phase flip in the $\delta-$correlated model
\bqn
|\psi\rr = |\phi\rr \otimes |\psi_1\rr \\
|\bar{\psi}\rr =   \lr \phi ,x|  U_{cpf} |\psi\rr \\
= \alpha|0\rr + \beta(\cos^2\theta - \sin^2\theta)|1\rr\\
\omega(\theta) = \lr \bar{\psi} | \bar{\psi}\rr = 1 - 4\beta^2(\sin^2\theta -\sin^4\theta)
\eqn
This gives us a partition function of
\bq
Z = 2\pi\int_0^\theta \omega(\theta) d\theta = 2\pi^2(1-\beta^2/2)
\eq
The density matrix for the circulating qubit will be
\bqn
\rho_{TM} = Z^{-1}\int_0^{2\pi}\int_0^\pi \omega(\theta)|\phi \rr\lr\phi|d\xi d\theta \\
= \int_0^{2\pi}\int_0^\pi\frac{1 - 4\beta^2(\sin^2\theta -\sin^4\theta)}{2\pi^2(1-\beta^2/2} \left( \begin{array}{cc} \cos^2\theta & e^{i\xi}\sin\theta\cos\theta \\\\ e^{-i\xi}\sin\theta\cos\theta & \sin^2\theta \end{array} \right)d\xi d\theta\\
=\frac{1}{2}I
\eqn
And for the control qubit we have
\bqn
\rho_{cpf} = \frac{2\pi}{Z}\int_0^\pi |\bar{\psi}\rr\lr\bar{\psi}|d\theta = \int_0^\pi\left( \begin{array}{cc} \alpha^2  & \alpha\beta(1-2\sin^2\theta) \\\\ \alpha\beta^\dagger(1 - 2\sin^2\theta) & \beta\beta^\dagger(1-2\sin^2\theta)^2 \end{array} \right)\frac{d\theta}{\pi(1-\beta^2/2)}\nb\\
=\left( \begin{array}{cc} 2\alpha^2/(\alpha^2+1)  & 0 \\\\ 0 & \beta\beta^\dagger/(\alpha^2+1) \end{array} \right)
\eqn
This particular decoherent time machine channel is half noise. Consequently the reduction in permitted states as well as the back-propagation effect onto the control qubit are both half strength, rather than complete elimination. The loss of off diagonal terms is due to the decoherent nature of the circulating bit.\\

\begin{figure}[h!]
  \caption{Phase flip version of the grandfather circuit.}
  \centering
    \includegraphics[width=0.5\textwidth]{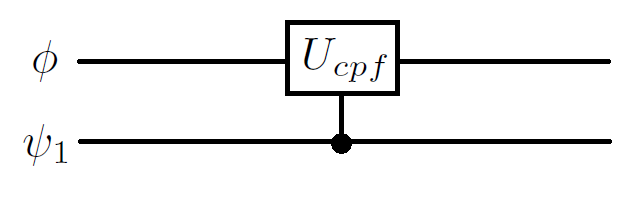}
\end{figure}
 In contrast to the previous examples, the phase flip operation does not have as significant an effect on the classical time machine channel. For the classical channel the weights are
\bqn
\lr 0,x | U_{cpf} | 0,\psi_1\rr = \alpha|0\rr + \beta|1\rr\\
\lr 1,x | U_{cpf} | 1,\psi_1\rr = \alpha|0\rr - \beta|1\rr\\
\omega_0 = k/2 + (1-k)| \lr 0,x | U_{cpf} | 0,\psi_1 \rr|^2= 1 - k/2\\
\omega_1 = k/2 + (1-k)| \lr 1,x | U_{cpf} | 1,\psi_1 \rr|^2 = 1 - k/2\\
Z= 2-k
\eqn
The partition function is independent of the control qubit, in contrast to both other cases. This is because the classical channel is only sensitive to the eigenvalues of the bit in its preferred decohering basis, and a the phase flip does not affect those values. 
The partition function for the time machine channel is
\bq 
\rho_{cl} = (2-k)^{-1}\left(\left (1-\frac{k}{2}\right)|0\rr\lr 0| + \left(1-\frac{k}{2}\right)|1\rr\lr 1|\right) = \frac{1}{2}I
\eq
and the control channel is given by
\bq
\rho_{clfp} = \frac{1}{2}\left( \alpha^2 |0\rr\lr 0|  + \beta\beta^\dagger |1\rr\lr 1| \right)
\eq
Again the interaction with a decoherent channel destroys the off diagonal terms. The effect is identical to measurement projection, with no need for back-propagation. The controlling bit becomes entangled with a forever hidden channel, making it too 'classical'. Circuits with more noise or less coherence seem to have less extreme effects than coherent noiseless ones. The requirement of noise or perturbation hints at more to be learned however. This brings us to the next circuit.


\section{Faulty gun circuit}

Taking a que from the perturbation of the grandfather scenario, let us consider general rotations and controlled rotations of the circulating qubit. Beginning with the Bell out state,
\bq
|\phi_{TM}\rr = \frac{1}{\sqrt{2}}( |00\rr + |11\rr)
\eq
We act with a rotation 
\bq
U_\zeta = \cos\zeta |0\rr\lr 0| + \sin\zeta |1\rr\lr 0| + \cos\zeta |1\rr\lr 1| - \sin\zeta |0\rr\lr 1|
\eq
onto the out channel, giving
\bq
|\psi'\rr = U_\zeta |\phi\rr = \frac{\cos\zeta}{\sqrt{2}} ( |00\rr + |11\rr ) + \frac{\sin\zeta}{\sqrt{2}}( |01\rr - |10\rr )
\eq 
The normalization constant is then given again by the inner product with the original Bell state,
\bq
N = \lr \phi_{TM} | U_\zeta | \phi_{TM} \rr = \cos\zeta
\eq
The parameter $\zeta$ rotates between the identity and the version of the grandfather circuit given by $U_{rot}$ in the previous section. The grandfather paradox occurs in the exact model when $N$ vanishes. 
\begin{figure}[h!]
  \caption{Measuring the effect of a single rotation on the periodic channel $\phi$.}
  \centering
    \includegraphics[width=0.5\textwidth]{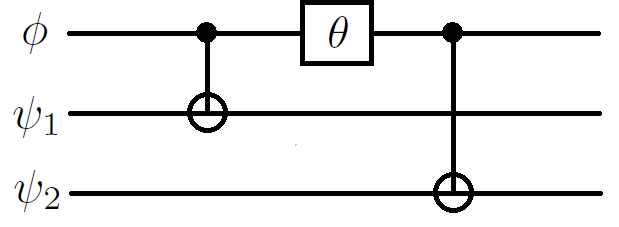}
\end{figure}
In the noisy classical limit we have a partition function of
\bq
Z = k + 2(1-k)\cos^2\zeta
\eq
This reflects the relative number of consistent histories available to the system. The $\delta-$correlated channel gives,
\bq
\omega = |\lr \phi | U_\zeta | \phi \rr |^2
\eq
and integrating over $|\phi\rr$,
\bq
|\phi\rr = \gamma_0|0\rr + \gamma_1 e^{i\xi}|1\rr = \cos\theta |0\rr + e^{i\xi}\sin\theta|1\rr
\eq
we have
\bq
\omega = | \cos\zeta - 2i \cos\theta \sin\theta \sin\zeta\sin\xi |^2
\eq
giving a partition function of
\bq
Z = \frac{\pi^2}{2}( 3\cos^2\zeta + 1 )
\eq
and a density matrix of
\bqn
\rho_{00} = Z^{-1} \int_0^\pi \int_0^{2\pi} \cos^2\theta [ \cos^2\zeta + 4\sin^2\zeta \sin^2 \xi \sin^2 \theta \cos^2 \theta ]d\theta d\xi =1/2\\
\rho_{11} = Z^{-1} \int_0^\pi \int_0^{2\pi} \sin^2\theta [ \cos^2\zeta + 4\sin^2\zeta \sin^2 \xi \sin^2 \theta \cos^2 \theta ]d\theta d\xi =1/2\\
\rho_{10}=\rho_{01} = Z^{-1} \int_0^\pi \int_0^{2\pi} e^{i\xi}\sin\theta\cos\theta [ \cos^2\zeta + 4\sin^2\zeta \sin^2 \xi \sin^2 \theta \cos^2 \theta ]d\theta d\xi =0
\eqn
The mixed state is mainly a function of the symmetry of the circuit, as well as its factorization. This effect will present itself in several other circuits as well. The role of the relative value of the partition function and normalization parameter as a sort of prior probability is more evident when we consider the controlled not version of the circuit.\\

Starting with an input control qubit defined as
\bq
|\psi_1\rr = \alpha|0\rr + \beta|1\rr
\eq
 we can add the maximally entangled pair of the in state and referce qubit.
\bq
|\phi_{out}\rr = \frac{1}{\sqrt{2}}( |00\rr + |11\rr)
\eq
The joint state is acted on by a controlled not 
\bq
U_{cn23} = |x,00\rr\lr x,00| + |x,01\rr\lr x,01| + |x,10\rr\lr x,11| + |x,11\rr\lr x,10|
\eq
giving an in state of
\bq
|\phi_{in}\rr = U_t|\phi_{out}\psi_1\rr = \frac{\alpha}{\sqrt{2}}( |000\rr + |110\rr) + \frac{\beta}{\sqrt{2}}( |011\rr + |101\rr)
\eq
projecting onto the pair, we get
\bq
\lr \phi_{out}^\dagger,x | U_t |\phi_{in}\rr = \alpha |0\rr = N|0\rr
\eq
The control qubit is forced into the state that avoids the paradox, regardless of the original amplitude or phase relative to the rest of the wave-function. Furthermore, the result holds even if we consider the control channel to be macroscopic. This is an indicator of how the nonlinear effects of state selection can propagate backwards into what we would normally consider the initial states of the system. Consider the action of $C_{not}$ from the outside channel onto a decoherent time machine channel.
\begin{figure}[h!]
  \caption{Controlled not acting on $\phi$.}
  \centering
    \includegraphics[width=0.3\textwidth]{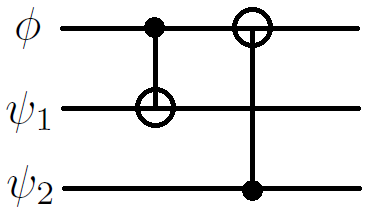}
\end{figure}
The $\delta$ model in the controlled not circuit gives,
\bqn
|\psi_1\rr = \alpha|0\rr + \beta|1\rr\\
|\phi_{out}\rr = \gamma_0 |0\rr + \gamma_1 |1\rr = \cos{\theta}|0\rr + e^{i\xi}\sin{\theta}|1\rr\\
|\psi\rr = \alpha \cos\theta |00\rr + \alpha e^{i\xi}\sin\theta |10\rr + \beta\cos\theta |01\rr + \beta e^{i\xi}\sin\theta |11\rr
\eqn
we get for the in state
\bq
|\phi_{in}\rr = U_t|\psi\rr = \alpha \cos\theta |00\rr + \alpha e^{i\xi}\sin\theta |10\rr + \beta\cos\theta |11\rr + \beta e^{i\xi}\sin\theta |01\rr
\eq
which projects to the state,
\bq
|\bar{\psi}\rr = \lr \phi_{out}^\dagger,x | U_t | \phi_{in} \rr = \alpha |0\rr + 2\beta \sin\theta \cos \theta \cos \xi |1\rr
\eq
the weight function is
\bq
N^2 = \omega(\theta,\xi) = | \lr \phi_{out}^\dagger,x | U_t | \phi_{in} \rr|^2 = \alpha^2 + 4\beta^2 \cos^2 \xi \sin^2\theta \cos^2\theta
\eq
and partition function of
\bq
Z = \int_0^\pi \int_0^{2\pi} N^2 d\theta d\xi = \frac{\pi^2}{2}( 3\alpha^2 + 1 )
\eq
and a density matrix given by
\bqn
\rho_{00} = \frac{2}{\pi^2}( 3\alpha^2 + 1 )^{-1}\int_0^\pi \int_0^{2\pi}  \omega(\theta,\xi)\cos^2\theta = \frac{1}{2}\\
\rho_{11} = \frac{2}{\pi^2}( 3\alpha^2 + 1 )^{-1}\int_0^\pi \int_0^{2\pi}  \omega(\theta,\xi)\sin^2\theta = \frac{1}{2}\\
\rho_{01} = \rho_{10} = 0
\eqn
We can also do an expectation value for the control bit, since the weight is also a function of $\alpha$ and $\beta$. The partition function $Z$ will serve as a weight that skews the statistics of a random mixture of control bits. In this case if we use $\zeta$ as the angle of the flip component of the control bit 
\bqn
|\psi_1\rr = \cos\zeta |0\rr + e^{i\vartheta}\sin\zeta |1\rr \\
\omega_{\zeta} = Z_{TM}(\alpha) = \frac{\pi^2}{2}(3\alpha^2 + 1) = \frac{\pi^2}{2}(3\cos^2\zeta + 1)
\eqn
A linear combination of density matrices weighted by $Z(\alpha)$ will give us
\bqn
Z_{\zeta} = 2\pi\int_0^{\pi} \left[ \frac{\pi^2}{2}(3\cos^2\zeta + 1) \right] d\zeta = \frac{5}{2}\pi^4 \\
\lr 0|\rho_\zeta|0\rr = \frac{2}{5}\pi^{-4} \int_0^{2\pi}d\vartheta \int_0^\pi d\zeta \left[\frac{\pi^2}{2}( 3\cos^4\zeta + \cos^2\zeta)\right] = \frac{13}{20}\\
\lr 1|\rho_\zeta|1\rr = \frac{2}{5}\pi^{-4} \int_0^{2\pi}d\vartheta \int_0^\pi d\zeta \left[\frac{\pi^2}{2}( 3\sin^2\zeta \cos^2\zeta + \sin^2\zeta)\right] = \frac{7}{20}\\
\lr 0|\rho_\zeta|1\rr = \frac{2}{5}\pi^{-4} \int_0^{2\pi}d\vartheta\int_0^\pi d\zeta \left[\frac{\pi^2}{2} e^{i\vartheta}( 3\sin\zeta\cos^3\zeta + \sin\zeta\cos\zeta)\right] = 0
\eqn
rather than the normal expectation value of $I/2$. This is an example of soft back-propagation of the constraints of the selection affecting normally flat distributions of randomly generated initial conditions. The incoming states are biased against creating the paradox. It is a fairly generic feature that the effects of time machines are felt outside of the respective light cones.\\
We can evaluate the circuit in the classical limit as
\bqn
Z_{cl} = k|\lr 0,x | U_t | 1,\psi_1\rr|^2  + k|\lr 1,x | U_t | 0,\psi_1\rr|^2+ (1-k)|\lr 0,x | U_t | 0,\psi_1\rr|^2 + (1-k)|\lr 1,x | U_t | 1,\psi_1\rr|^2 \nb\\
= 2k(1-\alpha^2) + 2(1-k)\alpha^2
\eqn
Due to symmetry the internal classical bit is mixed.
\bq
\rho_{cl} = \frac{1}{2}I
\eq
The expectation value of the control qubit in the classical time machine case is
\bqn
Z_\zeta = \int_0^{2\pi} \int_0^\pi (  2k\sin^2\zeta+ 2(1-k)\cos^2 \zeta ) d\vartheta d\zeta = 4\pi^2\\
\lr 0 | \rho_\zeta |0\rr = \pi^{-1} \int_0^\pi d\zeta (2k\sin^2\zeta\cos^2\zeta + 2(1-k)\cos^4\zeta ) = (3-2k)/8\\
\lr 1 | \rho_\zeta |1\rr = \pi^{-1} \int_0^\pi d\zeta (2k\sin^4\zeta + 2(1-k)\sin^2\zeta\cos^2\zeta ) = (1+2k)/8\\
\lr 0 | \rho_\zeta | 1 \rr= 0
\eqn
With $k$ ranging from 0 to 1. This is considering only the circulating bit to be classical. If we take the classical limit on both the control and the circulating bit we get,
\bqn
Z=2\\
\lr 0 | \rho_\zeta | 0\rr =  \frac{1}{2}(2-k)\\
\lr 1 | \rho_\zeta | 1 \rr = \frac{k}{2}
\eqn
The classical bit is forced up to the level of noise to avoid activating the $C_{not}$ gate. This is useful expression can be applied to determine approximate behavior of classical time machine circuits. 
Finally the standard depolarizing channel should be analyzed under the action of each singular circuit. Recall the definition of the four orthogonal projections for the Bell state prescription.
\bqn
|\bar{\psi}_B\rr = \lr 00 + 11 ,x| U_t | \phi_{Bell}, \psi_{ex} \rr\\
|\bar{\psi}_-\rr = \lr 00 - 11  ,x| U_t | \phi_{Bell}, \psi_{ex} \rr\\
|\bar{\psi}_N\rr = \lr 01 + 10 ,x| U_t | \phi_{Bell}, \psi_{ex} \rr\\
|\bar{\psi}_{N-}\rr \lr 01 - 10  ,x| U_t | \phi_{Bell}, \psi_{ex} \rr
\eqn
The partition function of the noisy Bell channel is,
\bq
Z_{\lambda} = (1-\lambda)\lr \bar{\psi}_B | \bar{\psi}_B\rr + \lambda/4
\eq
The relevent out states are all the same product of $\phi_{bell}$ and the control qubit $\psi_1$. The in states are given by the action of each circuit
\bqn
|\psi_{cpf-in}\rr =\frac{\alpha}{\sqrt{2}} (|00\rr + |11\rr)\otimes |0\rr + \frac{\beta}{\sqrt{2}} ( |00\rr - |11\rr ) \otimes |1\rr \\
|\psi_{crot-in}\rr =\frac{\alpha}{\sqrt{2}} (|00\rr + |11\rr)\otimes |0\rr + \frac{\beta}{\sqrt{2}} ( |01\rr - |10\rr ) \otimes |1\rr \\
|\psi_{cnot-in}\rr =\frac{\alpha}{\sqrt{2}} (|00\rr + |11\rr)\otimes |0\rr + \frac{\beta}{\sqrt{2}} ( |01\rr + |10\rr ) \otimes |1\rr 
\eqn
In each of these cases we have
\bq
|\bar{\psi}_B\rr = \alpha|0\rr
\eq
For each of the others, one other projection takes on a nonzero value. In the controlled phase flip circuit,
\bq
|\bar{\psi}_-\rr =\beta|1\rr
\eq
In the controlled not circuit,
\bq
|\bar{\psi}_N\rr = \beta|1\rr
\eq
In the controlled rotation circuit
\bq
|\bar{\psi}_{N-}\rr = \beta|1\rr
\eq
This gives rise to three identical density matrices and partition functions.
\bqn
Z_{cpf}=Z_{cnot}=Z_{crot}= (1-\lambda)\alpha^2 + \lambda/4\\
\rho_{cpf} = \rho_{cnot} = \rho_{crot} =  (1-\frac{3\lambda}{4})\frac{\alpha^2}{Z} |0\rr\lr 0| + \frac{\lambda\beta\beta^\dagger}{4Z}|1\rr\lr 1| 
\eqn
In the case of rotation through an arbitrary angle $\zeta$, or by an arbitrary phase $\xi$, the in states are given by
\bqn
|\psi_{\zeta-in}\rr =\frac{1}{\sqrt{2}}\left(|00\rr + |11\rr\right)\otimes\left(\alpha|0\rr + \beta\cos\zeta|1\rr\right) + \sin\zeta\frac{\beta}{\sqrt{2}} (|01\rr - |10\rr ) \otimes |1\rr \\
|\psi_{\xi-in}\rr =\frac{\alpha}{\sqrt{2}} (|00\rr + |11\rr)\otimes |0\rr + \frac{\beta}{\sqrt{2}} ( |00\rr + e^{i\xi}|11\rr ) \otimes |1\rr 
\eqn
\begin{figure}[h!]
  \caption{Controlled rotation of decoherent $\phi$.}
  \centering
    \includegraphics[width=0.3\textwidth]{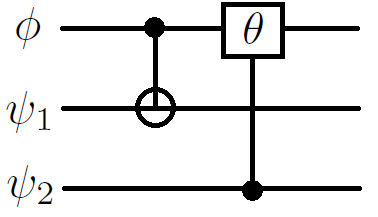}
\end{figure}
Two projections are nonzero for the $\zeta$ circuit.
\bqn
|\bar{\psi}_{\zeta,B}\rr = \alpha|0\rr + \beta\cos\zeta|1\rr\\
|\bar{\psi}_{\zeta,N-}\rr = \beta\sin\zeta|1\rr\\
N^2_B=\lr\bar{\psi}_{\zeta,B}|\bar{\psi}_{\zeta,B}\rr = 1-\beta\beta^\dagger\sin^2\zeta\\
N^2_{N-}=\lr\bar{\psi}_{\zeta,N-}|\bar{\psi}_{\zeta,N-}\rr = \beta\beta^\dagger\sin^2\zeta
\eqn
The partition function is
\bq
Z_\zeta = (1-\lambda)N^2_B + \lambda/4 = 1 - \frac{3}{4}\lambda - (1-\lambda)\beta\beta^\dagger\sin^2\zeta
\eq
and the projected density matrix is
\bqn
\rho_\zeta = \frac{\alpha^2}{Z}\left(1-\frac{3\lambda}{4}\right)|0\rr\lr 0| + \frac{\beta\beta^\dagger}{Z}\left( \frac{\lambda}{4} + (1-\lambda)\cos^2\zeta\right)|1\rr\lr 1|\nb\\
+\frac{\cos\zeta}{Z}\left(1 - \frac{3\lambda}{4}\right)\left( \alpha\beta |1\rr\lr 0| + \alpha\beta^\dagger |0\rr\lr 1| \right)
\eqn
Two projections are also nonzero for the $\xi$ circuit
\bqn
|\bar{\psi}_{\xi,B}\rr = \alpha|0\rr + \frac{\beta}{2}(1+e^{i\xi})|1\rr\\
|\bar{\psi}_{\xi,-}\rr = \frac{\beta}{2}(1-e^{i\xi})|1\rr\\
N^2_B= \lr\bar{\psi}_{\xi,B}|\bar{\psi}_{\xi,B}\rr =1 + \frac{\beta\beta^\dagger}{2}(1 + \cos \xi)\\
N^2_{-}=\lr\bar{\psi}_{\xi,-}|\bar{\psi}_{\xi,-}\rr = \frac{\beta\beta^\dagger}{2}(1-\cos\xi)
\eqn
with partition function
\bq
Z_\xi = (1-\lambda)N^2_B + \lambda/4 =1 - \frac{3}{4}\lambda + (1-\lambda)\frac{\beta\beta^\dagger}{2}(1 + \cos \xi)
\eq
and density matrix of
\bqn
\rho_\xi =\frac{\alpha^2}{Z}\left(1-\frac{3\lambda}{4}\right)|0\rr\lr 0| +\frac{\beta\beta^\dagger}{4Z}\left( \frac{\lambda}{4} + 2(1-\lambda)(1+\cos\xi)\right)|1\rr\lr 1|\nb\\
+\frac{4-3\lambda}{4Z} \left( \alpha\beta(1+e^{i\xi})|1\rr\lr 0| + \alpha\beta^\dagger(1+e^{-i\xi})|0\rr\lr 1| \right)
\eqn

We will return later to the full implications of these expressions. For now let us examine how time machines interact with measurement.


\section{Unproven proofs}

The idea of the unproven proof is that a circulating bit contains represents an inaccessible component of the system at any time outside of the loop. In the standard theory of entanglement mixed states are produced by the action of a trace over hidden degrees of freedom. The hidden degrees of freedom are called the purification space of the system. In scenarios with time machines, the degrees of freedom of a system appear change. The internal states of the time machine represent a temporary increase in the dimension of the system. If they become entangled and then later disappear, a sort of information paradox is born. The purification space as such ceases to exist, so what becomes of the rest of the wave-function? Does it remain mixed or pure? In the coherent approach, the state can be made pure, leading to strong acausal effects. These effects are less pronounced with noise, but at the cost of leaving a mixed state. The final state projection model is exactly the application of this method of purification to the state left after a black hole has evaporated\cite{mald}. This case has also appeared in \cite{seth3}, included here for completeness and discussion with other treatments. 
\begin{figure}[h!]
  \caption{The unproven proof circuit uses $\phi$ as the control qubit.}
  \centering
    \includegraphics[width=0.5\textwidth]{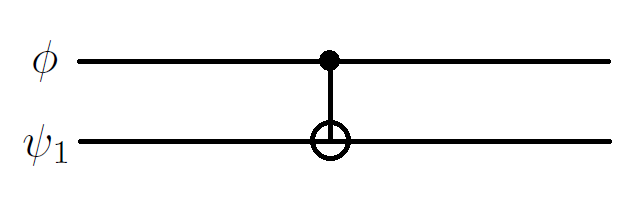}
\end{figure}
 The first circuit to consider here is a single external qubit acted on by  a $C_{not}$ controlled by the time machine state. A simple version of a measurement. We initialize the incoming state,
\bq
|\psi_1\rr = |0\rr
\eq
Then taking the product with the Bell pair state,
\bqn
|\phi_{Bell}\rr = \frac{1}{\sqrt{2}}( |00\rr + |11\rr)\\
|\psi_{out}\rr = |\phi_{Bell}\rr \otimes |\psi_1\rr = \frac{1}{\sqrt{2}}( |000\rr + |110\rr )
\eqn
 Now the measurement operation defined here as $C_{not}$ from qubit 2 to qubit 3 gives
\bq
|\psi_{in}\rr = U_{cnot23} |\psi_{out}\rr = \frac{1}{\sqrt{2}}(|000\rr + |111\rr) 
\eq
and the projection gives
\bq
|\bar{\psi}\rr = \lr \phi_{Bell} | \psi_{in}\rr = \frac{1}{2}(|0\rr + |1\rr)
\eq
This looks like a fairly benign result. The probing qubit is simply rotated, the resulting superposition no different than the action of an ordinary beam splitter. For a more generic incoming state we have
\bqn
|\psi_1\rr =\alpha |0\rr + \beta|1\rr \\
|\psi_{out}\rr = \frac{\alpha}{\sqrt{2}} ( |000\rr + |110\rr ) + \frac{\beta}{\sqrt{2}}( | 001\rr + |111\rr ) \\
|\psi_{in}\rr = U_{cnot23} |\psi_{out}\rr = \frac{\alpha}{\sqrt{2}} ( |000\rr + |111\rr ) + \frac{\beta}{\sqrt{2}}( | 001\rr + |110\rr )\\
|\bar{\psi}\rr = \lr \phi_{Bell} | \psi_{in}\rr = \frac{\alpha +\beta}2(|0\rr + |1\rr) 
\eqn
This is quite a bit more shocking. The state final state, while pure, is extremely non-unitary.  The relative amplitude and phase of the original state of the probe are absorbed into the normalization constant, giving an out state independent of the input. The circuit in this method 'forgets', making it an 'amnesia' type circuit. This many to one behavior is another indication that time machines or other post-selection type phenomena must be closely related to entropic phenomena if the second law of thermodynamics is to hold.  Just  like the grandfather paradox, an effective prior probability that back-propagates onto the initial state of $\psi_1$. The relative acceptance rate of the post-selection is 
\bq
N^2 = \frac{1}2(1 + \alpha\beta + \alpha\beta^\dagger) 
\eq
This is basically the probability that a measurement of the out qubit along the rotated axis of $|0\rr +|1\rr$ will find the qubit in that state. The system appears to be equivalent to using post-selection directly on the channel $\psi_1$, fixing its value as that superposition. Indeed there is a certain similarity between time machines and classical measurement apparatus, due to the familiar 'collapse postulate'. We will return to this and related scenarios in the amnesia circuit section.\\
In the noisy Bell model, one other orthogonal projection is not identically zero.
\bqn
|\bar{\psi}_B\rr = \lr 00 + 11 ,x| \psi_{in} \rr = \frac12(\alpha +\beta)(|0\rr + |1\rr) \\
|\bar{\psi}_-\rr = \lr 00 - 11  ,x| U_{cnot23} | \phi_{Bell}, \psi_{ex} \rr=\frac12 (\alpha -\beta)(|0\rr - |1\rr)
\eqn
Allowing the mixture to contain some portion of this second projection results in decoherence and is enough to avoid the measurement catastrophe.
The partition function is
\bq
Z_\lambda = (1-\lambda)\lr \bar{\psi}_B | \bar{\psi}_B\rr + \lambda/4 = \frac12 - \frac\lambda4  + \frac{1-\lambda}2(\alpha\beta + \alpha\beta^\dagger) 
\eq
This gives a density matrix for the probe channel of
\bqn
\rho_{00}= \rho_{11} = \frac1{4Z}\left(  (1-\lambda)|\alpha+\beta|^2 + \frac\lambda2 \right) = \frac12\\
\rho_{10}=\rho_{01}^\dagger =  \frac1{4Z}\left(  (1-\lambda)|\alpha+\beta|^2 + \frac\lambda2 \alpha(\beta+\beta^\dagger)\right) = \frac12 +\frac\lambda{4Z}(N_B^2-1)
\eqn
The correction to the off diagonal term indicates some distortion towards a preferred superposition state, but some decoherence as well. For the classical limit of the unproven proof circuit, the states are
\bqn
|0,\psi_1\rr = |0\rr \otimes |\psi_1\rr = \alpha|00\rr + \beta|01\rr\\
|1,\psi_1\rr = |1\rr \otimes |\psi_1\rr = \alpha|10\rr + \beta|11\rr\\
U_m|0,\psi_1\rr = |0,\psi_1\rr\\
U_m|1,\psi_1\rr = \alpha|11\rr + \beta|10\rr
\eqn

 The two weights for the time machine's classical internal states are
\bqn
\omega_0 = k| \lr 1,x|U_m|0,\psi_1\rr |^2 + (1-k)| \lr 0,x|U_m|0,\psi_1\rr |^2 = 1 - k\\
\omega_1 = k| \lr 0,x|U_m|1,\psi_1\rr |^2 + (1-k)| \lr 1,x|U_m|1,\psi_1\rr |^2  = 1 - k
\eqn
and the partition function is just
\bq
Z = \omega_0 + \omega_1  =  2-2k
\eq
giving a density matrix for the probe channel of
\bq
\rho_{final} = \frac{1-k}{Z}\left( |\psi_1\rr\lr \psi_1^\dagger| + U_{not}|\psi_1\rr\lr\psi_1^\dagger|U_{not}^\dagger \right) = \frac{1}{2}\left( \begin{array}{cc} 1 & \alpha\beta \\  \alpha\beta^\dagger & 1 \end{array} \right)
\eq
The probe bit is partially randomized by the briefly accessible random classical bit from the loop, but may retain coherence based on how close it is to the gate invariant state $|0\rr+|1\rr$. In this case the classical limit is better behaved than the Bell state model, since the wave function amplitude never vanishes, and the partition function is independent of the incoming state. Another alternative implementation of measurement is the conditional rotation gate controlled by the circulating qubit. 
\bq
U_{crot} = |00\rr\lr 00| + | 01\rr\lr 01| + |11\rr\lr 10| - |10\rr\lr 11|
\eq
And the conditional phase flip which is symmetric between the two channels.
\bq
U_{cpf} = |00\rr\lr 00| + |01\rr\lr 01| + |10\rr\lr 10| - |11\rr\lr 11|
\eq
Taking the product with the Bell state
\bq
|\psi_{out}\rr = |\psi_{TM}\rr\otimes |\psi_1\rr = \frac{1}{\sqrt{2}}\left( \alpha|000\rr + \alpha|110\rr + \beta|001\rr + \beta |111\rr\right)
\eq
The respective in states are
\bqn
|\psi_{crot-in}\rr = U_{crot} |\psi_{out}\rr = \frac{1}{\sqrt{2}}\left(  \alpha |000\rr +\alpha|111\rr + \beta |001\rr - \beta |110\rr\right)\\
|\psi_{cpf-in}\rr = U_{cpf} |\psi_{out}\rr = \frac{1}{\sqrt{2}}\left(  \alpha|000\rr + \alpha|110\rr + \beta|001\rr - \beta |111\rr \right)
\eqn
An these project to the respective final probe states
\bqn
|\bar{\psi}_{crot}\rr = \lr \phi_{TM} ,x|\psi_{crot-in}\rr = \frac{\alpha -\beta}{2} |0\rr + \frac{\alpha +\beta}{2}|1\rr \\
|\bar{\psi}_{cpf}\rr = \lr \phi_{TM},x|\psi_{cpf-in}\rr = \alpha |0\rr 
\eqn
For the controlled flip we still have an amnesia type circuit and a measurement catastrophe similar to the grandfather paradox when $\alpha$ vanishes. Without the ability to trace over $\phi$, destructive interference causes the wavefunction to vanish, and its renormalized direction to be indeterminate. 
The normalization factors are
\bqn
N^2_{crot} = 1/2\\
N_{cpf} = \alpha
\eqn
To study these circuits in the partially decoherent $\delta-$weight model , we begin with the standard 2-qubit product state
\bq
|\psi_{out}\rr = \alpha\cos\theta |00\rr + \beta\cos\theta|01\rr + \alpha e^{i\xi}\sin\theta|10\rr + \beta e^{i\xi}\sin\theta|11\rr
\eq
The action of $U_{cpf}$ is symmetric between the channels, and the analysis of it has already been covered a previous section. The action of $U_{crot}$ controlled by the circulating qubit, gives the state
\bq
|\psi_{in}\rr = \alpha\cos\theta |00\rr + \beta\cos\theta|01\rr  - \beta e^{i\xi}\sin\theta|10\rr + \alpha e^{i\xi}\sin\theta|11\rr
\eq
Then projecting against the initial time machine qubit,
\bq
|\bar{\psi}\rr = \lr \phi_{TM} ,x | \psi_{in}\rr = (\alpha \cos^2\theta - \beta\sin^2\theta) |0\rr + (\beta\cos^2\theta +\alpha\sin^2\theta)|1\rr
\eq
This gives us a weight of
\bq
\omega(\theta,\xi) =\lr \bar{\psi} | \bar{\psi}\rr =  \sin^4\theta + \cos^4\theta
\eq
and partition function of 
\bq
Z= 3\pi^2/2
\eq
The density matrix for $\phi$ is
\bq
\rho_{TM} = Z^{-1}\int_0^\pi\int_0^{2\pi} \omega(\theta,\xi) \left( \begin{array}{cc} \cos^2\theta & e^{i\xi}\sin\theta\cos\theta \\  e^{-i\xi}\sin\theta\cos\theta & \sin^2\theta \end{array} \right) d\theta d\xi = \frac{1}{2}I
\eq
and for the probe state,
\bqn
\rho_{cpf} = Z^{-1}\int_0^\pi\int_0^{2\pi} | \bar{\psi} \rr\lr \bar{\psi}|  d\theta d\xi \\
\rho_{00} = \frac{1}{2} - \frac{\alpha}{6}(\beta^\dagger + \beta)\\
\rho_{11} = \frac{1}{2} + \frac{\alpha}{6}(\beta^\dagger + \beta)\\
\rho_{01} = \frac{\alpha}{2}(\beta^\dagger -\beta) +\frac{1}{6}(\alpha^2 - \beta\beta^\dagger)
\eqn
The classical limit weight function on the other hand is the same as before since the gate leaves the periodic channel unchanged.
\bqn
\omega_0 = k| \lr 1,x|U_{crot}|0,\psi_1\rr |^2 + (1-k)| \lr 0,x|U_{crot}|0,\psi_1\rr |^2 = 1 - k\\
\omega_1 = k| \lr 0,x|U_{crot}|1,\psi_1\rr |^2 + (1-k)| \lr 1,x|U_{crot}|1,\psi_1\rr |^2  = 1 - k\\
Z = 2-2k
\eqn
The density matrix is
\bq
\rho_{TM} = Z^{-1} ( \omega_0 |0\rr\lr 0| + \omega_1 |1\rr\lr 1| ) = \frac{1}{2}I
\eq
for the time machine and
\bq
\rho_p = Z^{-1} \left( \omega_0 |\psi_1\rr\lr \psi_1 | + \omega_1 U_{crot}|\psi_1\rr\lr \psi_1|U_{crot}^\dagger \right) = \frac{1}{2}\left( \begin{array}{cc} 1 & \alpha(\beta^\dagger-\beta) \\  \alpha(\beta - \beta^\dagger) & 1 \end{array} \right)
\eq
for the probe channel.


\section{Twice watched pot}

In the modern understanding of decoherence, it can occur from interaction with a hidden subsystem of environment. In isolated systems, the decoherence of a particular qubit is simply the extent to which the bit is entangled with a large number of other observables. The classical states can be measured by testing any one of a large number of degrees of freedom.

In order to further explore the measurement catastrophe and grandfather paradoxes, multiple probe qubits may be used to decohere the system. Consider the unproven proof circuit using $C_{not}$ from the periodic qubit $\psi_{TM}$ onto a probe qubit. Let us add a second probe channel controlled either by $\phi$ or by the first probe's value after it measures $\phi$. \begin{figure}[h!]
  \caption{Two measurements of $\phi$.}
  \centering
    \includegraphics[width=0.4\textwidth]{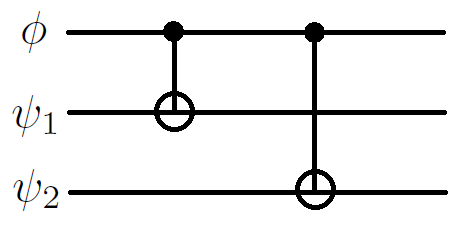}
\end{figure}
The standard out state in the Bell model gives
\bq
|\psi_{out}\rr = |\phi_{TM}\rr \otimes |\psi_1\rr \otimes |\psi_2\rr = \frac{1}{\sqrt{2}}( |0000\rr + |1100\rr)
\eq 
The action of either circuit will give the same in state
\bq
|\psi_{in}\rr =  \frac{1}{\sqrt{2}}( |0000\rr + |1111\rr)
\eq
which projects down to an ordinary Bell state , at least for inputs of $|0\rr$ for the probe channels.
\bq
|\bar{\psi}\rr = \lr \phi_{TM},xx|\psi_{in}\rr =  \frac{1}{2}( |00\rr + |11\rr)
\eq
Tracing over the second probe channel gives a full mixed state. Considering an arbitrary incoming product state for the probe qubits,
\bq
|\psi_{in}\rr =  |\phi_{TM}\rr \otimes |\psi_1\rr \otimes |\psi_2\rr = \frac{1}{\sqrt{2}}(|00\rr + |11\rr) \otimes (\alpha_1|0\rr + \beta_1|1\rr) \otimes (\alpha_2|0\rr + \beta_2|1\rr)
\eq
Acting with $C_{not}$ from $\phi$ onto both probes gives the in state
\bq
|\psi_{in}\rr = \frac{1}{\sqrt{2}} \left[ |00\rr \otimes  (\alpha_1|0\rr + \beta_1|1\rr) \otimes (\alpha_2|0\rr + \beta_2|1\rr) +  |11\rr \otimes  (\alpha_1|1\rr + \beta_1|0\rr) \otimes (\alpha_2|1\rr + \beta_2|0\rr) \right]
\eq
which projects down to
\bq
|\bar{\psi}_{Bell}\rr =  \lr \phi_{TM},xx|\psi_{in}\rr = \frac12(\alpha_1\alpha_2 + \beta_1\beta_2)(|00\rr + |11\rr)  + \frac12 (\alpha_1\beta_2 + \alpha_2\beta_1)(|10\rr + |01\rr)
\eq
This gives us a normalization factor of
\bq
N^2 = \lr \bar{\psi}_{Bell}|\bar{\psi}_{Bell}\rr = 1 + \alpha_1\alpha_2(\beta_1 + \beta_1^\dagger)(\beta_2 + \beta_2^\dagger)
\eq
This gives us two singular joint states, so the measurement catastrophe remains, however the final joint state is no longer independent of the initial probe values. The result should be quite surprising given the single probe unproven proof circuit result. We can attempt to apply the projection at different times, but the result is the same. The information loss creeps into all channels entangled with an erased channel.
\begin{figure}[h!]
  \caption{Projection affects channels that do not directly interact with $\phi$.}
  \centering
    \includegraphics[width=0.3\textwidth]{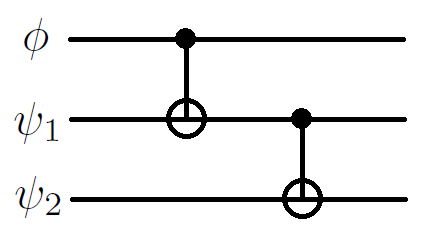}
\end{figure}

  If we separate the two circuits, considering the first measurement of $\phi$ by $\psi_1$ as a projective circuit, and then the later measurement of $\psi_1$ by $\psi_2$ as a normal unitary quantum gate, we would expect a joint state before the second measurement but after the time loop of
\bq
|\bar{\psi}\rr = \lr \phi_{TM},x | \psi_{in}\rr = \frac{1}{\sqrt{2}} (|0\rr + |1\rr) \otimes |\psi_2\rr 
\eq
The second channel then interacts after projection to give 
\bqn
U_{cn12}|\bar{\psi}\rr = \frac{1}{\sqrt{2}}( \alpha_2|00\rr + \beta_2|01\rr + \beta_2|10\rr + \alpha_2|11\rr)\nb\\
=\frac{\alpha_2}{\sqrt{2}}(|00\rr +|11\rr) + \frac{\beta_2}{\sqrt{2}}(|01\rr +|10\rr) 
\eqn
Carrying out the $U_{cn12}$ gate before projection gives the state as
\bqn
|\psi_{in}\rr = \alpha_1\alpha_2|0000\rr + \alpha_1\beta_2|0001\rr + \beta_1\beta_2|0010\rr + \beta_1\alpha_2|0011\rr\nb\\
+\beta_1\alpha_2|1100\rr+\beta_1\beta_2|1101\rr + \alpha_1\beta_2|1101\rr + \alpha_1\alpha_2|1111\rr
\eqn
Which then projects down to the previous result.
Returning to the simple double probe circuit,we wil consider the noisy case.  The nonzero projections of the probe product states are
\bqn
|\bar{\psi}_B\rr =  \frac12(\alpha_1\alpha_2 + \beta_1\beta_2)(|00\rr + |11\rr)  + \frac12 (\alpha_1\beta_2 + \alpha_2\beta_1)(|10\rr + |01\rr)\\
|\bar{\psi}_-\rr =  \frac12(\alpha_1\alpha_2 - \beta_1\beta_2)(|00\rr - |11\rr)  + \frac12 (\alpha_1\beta_2 - \alpha_2\beta_1)(|01\rr - |10\rr)
\eqn
The partition function is
\bq
Z_\lambda = (1-\lambda)N_B^2 + \lambda/4 = \frac12 - \frac14\lambda + \frac12(1-\lambda)\alpha_1\alpha_2(\beta_1+\beta_1^\dagger)(\beta_2+\beta_2^\dagger)
\eq
and the density matrix is
\bqn
\rho_\lambda = Z_\lambda^{-1}\left(1-\frac{3\lambda}4\right)|\bar{\psi}_B\rr\lr \bar{\psi}_B| + \frac\lambda{4Z_\lambda}|\bar{\psi}_-\rr\lr\bar{\psi}_-|\nb\\
\eqn

A few simple cases should help to see the physical picture here. For an initial probe value of $|00\rr$ we have
\bqn
\alpha_1=\alpha_2=1\\
Z =\frac12 - \frac\lambda4\\
\lr 00 | \rho_{twp} | 00\rr = \lr 11 | \rho_{twp} | 11\rr = 1/2
\eqn
with all other $\rho_{ij}$ vanishing. $Z$ has a local minimum at the two points where the projection against the Bell state vanishes. Instead of the measurement catastrophe, we have a factorable density matrix and pure out state 
\bqn
\alpha_1=\alpha_2=\beta_1 = -\beta_2 = 1/\sqrt{2}\\
Z =\lambda/4\\
|\xi_{min}\rr =|\bar{\psi}_-\rr =  \frac{1}{2} \left( |00\rr - |01\rr + |10\rr - |11\rr \right)\\
\rho_{min} = |\xi_{min}\rr\lr \xi_{min}|
\eqn
An orthogonal pure state gives the maximum value for $Z$ 
\bqn
\alpha_1=\alpha_2=\beta_1 = \beta_2 = 1/\sqrt{2}\\
Z = 1 - \frac34\lambda \\
|\xi_{max}\rr =|\bar{\psi}_B\rr=  \frac{1}{2} \left( |00\rr + |01\rr + |10\rr + |11\rr \right)\\
\rho_{max} = |\xi_{max}\rr\lr \xi_{max}|
\eqn
To get the classical limit for the double measurement circuit we take the classical weights defined as the amplitude squared of the projection against each classical basis state.
\bqn
|\psi_{out}(i)\rr = |e_i\rr \otimes |\psi_1\rr \otimes |\psi_2\rr\\
|\bar{\psi}_{ij}\rr = \lr e_i ,xx | U_c | \psi_{out}(j) \rr\\
|\bar{\psi}_{00}\rr = |\psi_1\rr \otimes |\psi_2\rr\\
|\bar{\psi}_{11}\rr = U_{not}|\psi_1\rr \otimes  U_{not} | \psi_2\rr \\
|\bar{\psi}_{01}\rr = |\bar{\psi}_{10}\rr = 0\\
\omega_0 = k\lr \bar{\psi}_{10} | \bar{\psi}_{10} \rr + (1-k)\lr \bar{\psi}_{00} | \bar{\psi}_{00} \rr = 1-k \\
\omega_1 = k\lr \bar{\psi}_{01} | \bar{\psi}_{01} \rr + (1-k)\lr \bar{\psi}_{11} | \bar{\psi}_{11} \rr = 1-k \\
Z = \omega_0 + \omega_1 = 2-2k
\eqn
This leads to the density matrix of the outgoing states
\bq
\rho_{cl} = \frac{1}{2}|\bar{\psi}_0\rr\lr\bar{\psi}_0| + \frac{1}{2}|\bar{\psi}_1\rr\lr\bar{\psi}_1| 
\eq
If the incoming channels are initialized to $|00\rr$ then this produces 
\bq
\rho = \frac{1}{2}( |00\rr\lr 00| + |11\rr\lr 11|)
\eq

Another interesting case is to consider the same circuit, but with the probe qubits initially in an entangled rather than product state.
\bq
|\psi_{out}\rr = |\phi_{Bell}\rr \otimes \left( \gamma_{00}|00\rr + \gamma_{01}|01\rr + \gamma_{10}|10\rr + \gamma_{11}|11\rr \right)
\eq
which projects to
\bqn
\lr \phi_{Bell},xx| \psi_{in}\rr = \lr \phi_{Bell},xx| \psi_{out}\rr + \lr \phi_{Bell},xx| U_{not(1)} U_{not(2)} | \psi_{out}\rr\\
|\bar{\psi}_B\rr=\frac{1}{2}(\gamma_{00}+\gamma_{11})\left( |00\rr +|11\rr \right) +\frac{1}{2}(\gamma_{01}+\gamma_{10})\left( |01\rr +|10\rr\right)
\eqn
which has a measurement catastrophe over the set of entangled initial probe states given by
\bq
|\psi_{null}\rr  =\cos\vartheta(|00\rr - |11\rr) + e^{i\xi}\sin\vartheta( |01\rr - |10\rr )
\eq
The dimension of the null space for the incoming measurement qubits remains at half, or one qubit less than, the whole space, as the number of measurement bits increases.
When adding noise there is again only one other nonzero projection.
\bq
|\bar{\psi}_-\rr = \frac12(\gamma_{00}-\gamma_{11})(|00\rr - |11\rr) + \frac12(\gamma_{01} - \gamma_{10})(|01\rr - |10\rr)
\eq
The partition function for the generic probe state is
\bq
Z_\lambda = (1-\lambda)N_B^2 + \lambda/4 = \frac12 - \frac\lambda4 + (1-\lambda)\left(\gamma_{00}\gamma_{11}^\dagger + \gamma_{11}\gamma_{00}^\dagger + \gamma_{01}\gamma_{10}^\dagger + \gamma_{10}\gamma_{01}^\dagger\right)
\eq
This is comparatively well behaved even for entangled in states, due to the effective noise contributed from the spread of the weight function. In the classical limit case the weights are unchanged by the addition of entanglement in the probe in states.\\
A second important point about decoherence is that it is not usually periodic. Entanglement between the circulating bits and the environment cannot generally be identical for in and out states. Consequently, if we have multiple CTC qubits, they cannot mutually decohere without the normalization parameter vanishing. Measurement circuits that entangle multiple time machine channels act similarly to grandfather circuits. For example a single $C_{not}$ connecting two periodic qubits immediately reduces the allowed states by half, as it fixes the value of the control qubit to avoid generating a paradox on the subject bit. The same qubit acting as a control for multiple periodic channels can effectively over-determine its state, creating a grandfather type paradox for all values of the incoming probe state. To analyze these circuits we will need two Bell states, $\phi_1$ and $\phi_2$ one for each periodic channel, and an input state $\psi_1$.  First let us examine the interaction of two periodic channels via $C_{not}$.
\begin{figure}[h!]
  \caption{Interaction between projected channels decreases the normalization factor.}
  \centering
    \includegraphics[width=0.4\textwidth]{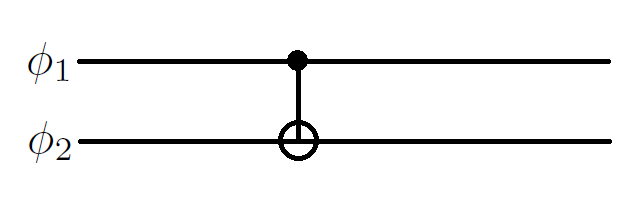}
\end{figure}
 Beginning with the out state in the Bell model
\bq
|\phi_{TM}\rr = |\phi_1\rr \otimes |\phi_2\rr = \frac{1}{2}\left( |0000\rr + |0011\rr + |1100\rr + |1111\rr \right)
\eq
We connect the two channels with a $C_{not}$ from bit 2 to bit 4. Bit 1 and bit 3 are the reference bits for 2 and 4 respectively.
\bq
|\phi_{in}\rr = U_{cn24}|\phi_{TM}\rr = \frac{1}{2}\left( |0000\rr + |0011\rr + |1101\rr + |1110\rr \right)
\eq
If we add a probe as bit 5, which measures bit 2 via another $C_{not}$ gate $U_{cn25}$, then the projected final state of the probe is unchanged. This indicates that channel $\phi_1$ consisting of bits 1 and 2 is fixed as the state $|00\rr$.

\bqn
|\psi_{out}\rr = |\phi_{TM}\rr \otimes |\psi_1\rr\\
|\psi_{in}\rr = U_{cn25}U_{cn24}|\psi_{out}\rr\nb\\
= \frac{1}{2}\left( |0000\rr + |0011\rr\right) \otimes |\psi_1\rr + \frac{1}{2}\left(  |1101\rr + |1110\rr \right) \otimes U_{not}|\psi_1\rr\\
|\bar{\psi}_B \rr = \lr \phi_2 , x | \lr \phi_1 ,xxx | \psi_{in}'\rr = \frac{1}{4}|\psi_1\rr
\eqn

Adding noise $\lambda$ requires us to use more projection states to cover all the degrees of freedom of the 4 qubit channel. 
There are potentially 16 independent projections given by the products of pairs of the four single channel orthogonal vectors.
In modeling the noise we have the freedom to correlate the fluctuations between channels $\phi_1$ and $\phi_2$ or not.
Four of the projections are nonzero.
\bqn
|\bar{\psi}_{BB}\rr = \lr \phi_{BB} | \psi_{in} \rr = \frac12|\psi_1\rr\\
|\bar{\psi}_{-B}\rr = \lr \phi_{-B} | \psi_{in} \rr = \frac12|\psi_1\rr\\
|\bar{\psi}_{BN}\rr = \lr \phi_{BN} | \psi_{in} \rr = \frac12U_{not}|\psi_1\rr\\
|\bar{\psi}_{-N}\rr = \lr \phi_{-N} | \psi_{in} \rr = - \frac12U_{not}|\psi_1\rr
\eqn
Weighting each projection by the number of independent errors, and assuming the error rate is $\lambda$ for each qubit, we get three terms.
\bqn
Z = (1-\frac34\lambda)^2\lr\bar{\psi}_{BB} |\bar{\psi}_{BB}\rr + \frac\lambda4(1-\frac34\lambda)\left( \lr\bar{\psi}_{-B} |\bar{\psi}_{-B}\rr+ \lr\bar{\psi}_{BN} |\bar{\psi}_{BN}\rr \right) + \frac{\lambda^2}{16} \lr\bar{\psi}_{-N} |\bar{\psi}_{-N}\rr\nb\\
=\frac14(1-\frac\lambda2)^2
\eqn
This gives a density matrix for the outgoing probe of
\bqn
\rho  =  Z^{-1}\left((1-\frac34\lambda)^2|\bar{\psi}_{BB} \rr\lr\bar{\psi}_{BB}| + \frac\lambda4(1-\frac34\lambda)\left( |\bar{\psi}_{-B} \rr\lr\bar{\psi}_{-B}|+ |\bar{\psi}_{BN} \rr\lr\bar{\psi}_{BN}| \right) + \frac{\lambda^2}{16}| \bar{\psi}_{-N} \rr\lr\bar{\psi}_{-N}|\right)\nb\\
=\frac{4-3\lambda}{4-2\lambda}|\psi_1\rr\lr\psi_1| + \frac\lambda{4-2\lambda}U_{not}|\psi_1\rr\lr\psi_1|U_{not}^\dagger
\eqn
Evaluating the classical limit the weights are simpler.
\bqn
\omega_{00} =\omega_{01} = k/4 - (1-k)| \lr 00,x | U_{cn12}U_{cn13} | 00,\psi_1\rr |^2 = 1 - 3k/4\\
\omega_{10} = \omega_{11} = k/4 - (1-k)| \lr 10,x | U_{cn12}U_{cn13} | 10,\psi_1\rr |^2 = k/4\\
Z = 2 - k
\eqn
and measurement output of
\bq
\rho_{cl} = (1-\frac{k}{2Z})|\psi_1\rr\lr \psi_1| + \frac{k}{2Z}U_{not}|\psi_1\rr\lr \psi_1|U_{not}^\dagger
\eq


\section{Mutual paradoxes}

To further explore the concept of over-determination of entanglement, we examine a variation of the grandfather paradox, pitting one periodic qubit against another. 
Two periodic channels $\phi_1$ and $\phi_2$ are both acted on by a controlled not from a single external channel $\psi_1$. in between these interactions a not gate acts on the $\psi_1$ channel.
In the bell state model the out state is
\bq
|\psi_{out}\rr = \frac{1}{2}\left( |00\rr + |11\rr \right) \otimes \left( |00\rr + |11\rr \right) \otimes \left( \alpha|0\rr + \beta|1\rr\right)
\eq
The action of three cages determines the in state to be
\bqn
|\psi_{in} \rr= U_{cn54}U_{not5}U_{cn52}|\psi_{out}\rr = \alpha\left( |00011\rr + |00101\rr + |11011\rr + |11101\rr\right) +\nb\\
\beta\left( |01000\rr + |01110\rr + |10000\rr + |10110\rr\right) 
\eqn
This state is always orthogonal to the original product state for any value of the input state $\psi_1$. 
If the not gate is replaced by a rotation of angle $\zeta$ then the paradox is averted and the normalization factor behaves the same
as it does in the faulty gun circuit. 
\bqn
|\psi_{in}'(\zeta)\rr = U_{cn54}U_{rot5}(\zeta)U_{cn52}|\psi_{out}\rr = \cos\zeta   U_{cn54}U_{cn52}|\psi_{out}\rr - \sin\zeta |\psi_{in}\rr\\
\lr \phi_1,x | \lr \phi_2,xxx | \psi_{in}'(\zeta) \rr = \cos\zeta  \lr \phi_1,x | \lr \phi_2,xxx | U_{cn54}U_{cn52}|\psi_{out}\rr =\alpha\cos\zeta |0\rr\\
N_B = \alpha\cos\zeta 
\eqn
The first controlled not gate removes the $|1\rr$ part of $\psi_1$, and the second does the same for the superposition created by the rotation gate.
\begin{figure}[h!]
  \caption{Mutual paradox due to two inconsistent post-selections.}
  \centering
    \includegraphics[width=0.5\textwidth]{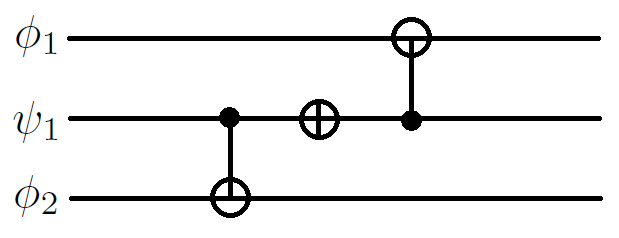}
\end{figure}

 The two loops do not have to be in causal contact with each other, the paradox is a property of the global structure of the circuit.
Adding noise again requires that we calculate the other nonzero projections of $|\psi_{in}'\rr$. These are
\bqn
|\bar{\psi}_{NB}\rr = -\beta\sin\zeta |0\rr\\
|\bar{\psi}_{BN}\rr = -\alpha\sin\zeta |1\rr\\
|\bar{\psi}_{NN}\rr =\beta\cos\zeta|1\rr
\eqn
Assuming independent noise of $\lambda$ in each periodic channel, we again have three terms in the partition function.
\bq
Z_\lambda(\alpha) = \left(1-\frac34\lambda\right)^2\alpha^2\cos^2\zeta + \frac\lambda4\left(1-\frac34\lambda\right)\sin^2\zeta + \frac{\lambda^2\beta\beta^\dagger}{16}\cos^2\zeta
\eq
The interaction with the decoherent channels mixes the initially pure state $\psi_1$ into
\bqn
\rho_\psi =  Z_\lambda^{-1}\left((1-\frac34\lambda)^2|\bar{\psi}_{BB} \rr\lr\bar{\psi}_{BB}| + \frac\lambda4(1-\frac34\lambda)\left( |\bar{\psi}_{NB} \rr\lr\bar{\psi}_{NB}|+ |\bar{\psi}_{BN} \rr\lr\bar{\psi}_{BN}| \right) + \frac{\lambda^2}{16}| \bar{\psi}_{NN} \rr\lr\bar{\psi}_{NN}|\right)\\
\rho_{00} =\frac1{Z_\lambda} \left(1-\frac34\lambda\right)^2\alpha^2\cos^2\zeta + \frac\lambda{4Z_\lambda}\left(1-\frac34\lambda\right)\beta\beta^\dagger\sin^2\zeta \\
\rho_{01}=0
\eqn
The circuit also has back-propagation, due to the dependence of $Z_\lambda$ on $\alpha$. The skew of an otherwise maximally mixed control channel can be calculated using
\bq
\rho_{00} = \frac{ \int \rho(\alpha) Z(\alpha) d\psi_1 }{\int Z d\psi_1}= \frac1{Z_\rho}\int_0^\pi\int_0^{2\pi} Z_\lambda(\cos\vartheta,\xi) \rho_\lambda(\vartheta,\xi) d\vartheta d\xi
\eq
where
\bqn
\alpha = \cos\vartheta\\
\beta = e^{i\xi}\sin\vartheta
\eqn
Evaluating the denominator gives
\bq
Z_\rho = \int Z_\lambda(\cos\theta) d\theta = \pi^2\left(1-\frac34\lambda\right)^2\cos^2\zeta + \pi^2\frac\lambda2\left(1-\frac34\lambda\right)\sin^2\zeta + \pi^2\frac{\lambda^2}{16}\cos^2\zeta
\eq
So an otherwise maximally mixed test channel $\psi_1$ will be distorted by post-selection into the mixed state
\bq
\bar{\rho}_{skew} = \frac{\pi^2}{Z_\rho} \left(1-\frac34\lambda\right)^2\cos^2\zeta + \frac{\lambda\pi^2}{4Z_\rho}\left(1-\frac34\lambda\right)\sin^2\zeta \\
\eq
The classical limit similar
\bqn
\lr 00,x| U_t | 00,\psi_1\rr =\lr 01,x| U_t | 01,\psi_1\rr =\lr 10,x| U_t | 10,\psi_1\rr =\lr 11,x| U_t | 11,\psi_1\rr = \alpha\cos\zeta\\
\lr 01,x| U_t | 00,\psi_1\rr =\lr 00,x| U_t | 01,\psi_1\rr =\lr 10,x| U_t | 11,\psi_1\rr =\lr 11,x| U_t | 10,\psi_1\rr = -\beta\sin\zeta\\
\lr 00,x| U_t | 10,\psi_1\rr =\lr 10,x| U_t | 00,\psi_1\rr =\lr 11,x| U_t | 01,\psi_1\rr =\lr 01,x| U_t | 11,\psi_1\rr = \alpha\sin\zeta\\
\lr 00,x| U_t | 11,\psi_1\rr =\lr 11,x| U_t | 00,\psi_1\rr =\lr 10,x| U_t | 01,\psi_1\rr =\lr 01,x| U_t | 10,\psi_1\rr = \beta\cos\zeta\\
Z_{cl}(\alpha) = (1-k)^2\alpha^2\cos^2\zeta + k(1-k)\sin^2\zeta + k^2\beta\beta^\dagger\cos^2\zeta
\eqn
carrying out the same calculation for skewing the input channel $\psi_1$ gives
\bq
Z_\rho = 2\pi\int_0^\pi Z_\alpha(\cos\theta) d\theta = \pi^2 \cos^2\zeta\left[ (1-k)^2 + k^2 \right] + 2\pi^2k(1-k)\sin^2\zeta 
\eq
The accessible states gives a skew onto the input qubit of
\bqn
\rho_{cl-\psi} = \frac{1}{Z_\rho}\int_0^\pi\int_0^{2\pi} Z_\alpha(\cos\theta)   \left( \begin{array}{cc} \cos^2\theta & e^{i\xi}\sin\theta\cos\theta \\  e^{-i\xi}\sin\theta\cos\theta & \sin^2\theta \end{array} \right) d\theta d\xi\nb\\
=\frac{\pi^2}{Z_\rho}k(1-k)\sin^2\zeta\left( \begin{array}{cc} 1 & 0 \\ 0 & 1 \end{array} \right) + \frac{\pi^2}{4Z_\alpha}\cos^2\zeta \left( \begin{array}{cc} 3(1-k)^2+k^2 & 0 \\ 0 & (1-k)^2 + 3k^2 \end{array} \right)
\eqn
The greater the noise, the more unbiased the input $\psi_1$. Input bias will be further explored in the section on back-propagation. 

\section{Third party paradox}

A second important example paradox spread out over multiple gates is the exclusive or grandfather circuit, 
consisting of two independent input channels and a single periodic channel which is acted on by both inputs acting as controls to two
$C_{not}$ gates. Projection over the periodic channel creates a correlation between the two input channels. The scenario is 
essentially the projection of a GHZ entanglement of three channels by the 'third party', the periodic channel, to create 
a Bell pair without direct interaction between the input qubits.  \\
The initial state in the Bell state description is 
\bqn
|\psi_{out}\rr = |\phi_{Bell} \rr \otimes |\psi_1\rr \otimes |\psi_2\rr\nb\\
 = \frac{1}{\sqrt{2}}\left( \alpha_1\alpha_2|0000\rr + \alpha_1\alpha_2|1100\rr  +\beta_1\beta_2|0011\rr + \beta_1\beta_2|1111\rr \right)\nb\\
+\frac{1}{\sqrt{2}}\left(\alpha_1\beta_2|0001\rr + \alpha_1\beta_2|1101\rr +\beta_1\alpha_2|0010\rr + \beta_1\alpha_2|1110\rr\right)
\eqn

The circuit consists of two controlled not gates giving,
\bqn
|\psi_{in}\rr = U_{cn32}U_{cn42}|\psi_{out}\rr \nb\\
= \frac{1}{\sqrt{2}}\left( \alpha_1\alpha_2|0000\rr + \alpha_1\alpha_2|1100\rr  +\beta_1\beta_2|0011\rr + \beta_1\beta_2|1111\rr \right)\nb\\
+\frac{1}{\sqrt{2}}\left(\alpha_1\beta_2|0001\rr + \alpha_1\beta_2|1101\rr +\beta_1\alpha_2|0010\rr + \beta_1\alpha_2|1110\rr\right)
\eqn

\begin{figure}[h!]
  \caption{Third party paradox produces entanglement between $\psi_1$ and $\psi_2$.}
  \centering
    \includegraphics[width=0.5\textwidth]{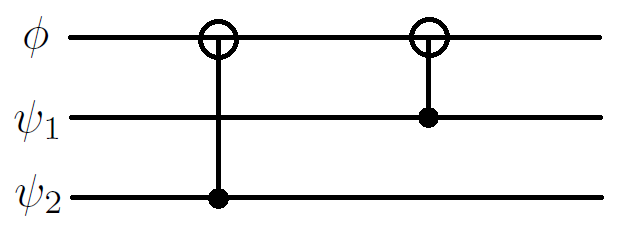}
\end{figure}

The projection of this state against $|\phi_{Bell}\rr$ gives
\bq
|\bar{\psi}_B\rr = \lr\phi_{bell}, xx|\psi_{in}\rr = \alpha_1\alpha_2|00\rr + \beta_1\beta_2|11\rr 
\eq
with a normalization factor
\bq
N^2 = \lr\bar{\psi}|\bar{\psi}\rr = \alpha_1^2\alpha_2^2 + \beta_1\beta_1^\dagger\beta_2\beta_2^\dagger
\eq
The product state is projected onto an entangled state. 
This is the mechanism most cited in information paradox literature for allowing information to escape and averting the firewall paradox in the black hole final state model.\cite{unit}
The singular states all correspond to orthogonal control bit inputs. 
\bq
N^2 = |\lr \psi_1 | \psi_2 \rr|^2
\eq
This is another indication of the global nature of paradoxes, and of the phenomenon of back-propagation. 
The noisy case adds only one other nonzero projection.
\bq
|\bar{\psi}_N\rr = \alpha_1\beta_2|01\rr + \alpha_2\beta_1|10\rr
\eq
The partition function and density matrix are
\bqn
Z_\lambda = (1-\lambda)N^2 + \lambda/4\\
\rho = \frac1{Z_\lambda}(1-\frac34\lambda)|\bar{\psi}_B\rr\lr\bar{\psi}_B| + \frac\lambda{4Z_\lambda}|\bar{\psi}_N\rr\lr \bar{\psi}_N|
\eqn
The effect is similar in the classical limit. The out states are given by 
\bqn
|\psi_{out}(0)\rr =  |0 \rr \otimes |\psi_1\rr \otimes |\psi_2\rr\nb\\
=\alpha_1\alpha_2|000\rr + \beta_1\beta_2|011\rr + \alpha_1\beta_2|001\rr +\beta_1\alpha_2|010\rr\\
|\psi_{out}(1)\rr =  |1 \rr \otimes |\psi_1\rr \otimes |\psi_2\rr\nb\\
=\alpha_1\alpha_2|100\rr + \beta_1\beta_2|111\rr + \alpha_1\beta_2|101\rr +\beta_1\alpha_2|110\rr
\eqn
and in states by
\bqn
|\psi_{in}(0)\rr = U_{cn31}U_{cn21}|\psi_{out}(0)\rr \nb\\
=\alpha_1\alpha_2|000\rr + \beta_1\beta_2|011\rr + \alpha_1\beta_2|101\rr +\beta_1\alpha_2|110\rr\\
|\psi_{in}(1)\rr = U_{cn31}U_{cn21}|\psi_{out}(1)\rr \nb\\
=\alpha_1\alpha_2|100\rr + \beta_1\beta_2|111\rr + \alpha_1\beta_2|001\rr +\beta_1\alpha_2|010\rr
\eqn
Then project each against the corresponding eigenstate to give
\bqn
|\bar{\psi}(0,0)\rr =|\bar{\psi}(1,1)\rr =  \lr 0, xx|\psi_{in}(0)\rr = |\bar{\psi}_B\rr \\
|\bar{\psi}(1,0)\rr =|\bar{\psi}(0,1)\rr = \lr 1, xx|\psi_{in}(1)\rr = |\bar{\psi}_N\rr 
\eqn
The resulting weights are then
\bqn
\omega_0 = (1-k)\lr \bar{\psi}(0,0) | \bar{\psi}(0,0) \rr + k\lr \bar{\psi}(1,0) | \bar{\psi}(1,0)\rr = (1-k)N^2 + k(1-N^2)\\
= \omega_1 = (1-k) \lr \bar{\psi}(1,1) | \bar{\psi}(1,1) \rr + k\lr \bar{\psi}(0,1) | \bar{\psi}(0,1)\rr\\
Z_{cl} = 2\omega_0
\eqn
And the density matrix as
\bq
\rho_{cl} =  \frac{1-k}{Z_{cl}}|\bar{\psi}_B\rr\lr\bar{\psi}_B| + \frac{k}{Z_{cl}}|\bar{\psi}_N\rr\lr \bar{\psi}_N|
\eq
If we fix one of the two controlling channels into a definite eigenstate, then its twin takes on the same state in the projected ensemble. 
The channels could be said to be 'super-entangled', since changes to the doublet state will still project onto another doublet.

%
%
\section{Stubborn spin effect}

There are several circuit representations of what has been called the unproven theorem paradox. The simplest model is a single qubit circulating along a CTC, which we then measure, effectively acting on it with a projection operator, obtaining classical information. The use of $P_\pm$ for measurement here means that we have effectively absorbed the tracing over external bits step into the normal evolution.
\bq
\alpha|+\rr + \beta|-\rr = |\phi\rr \rightarrow P_\pm|\phi\rr 
\eq
Just after a measurement, the system is in one of the two eigenstates. To compare the relative frequency of the two measurement outcomes, $|+\rr$, and $|-\rr$, we consider the weights. Both are eigenvectors of $P_\pm$, and the noise operator is symmetric between them, so the weights will be the same. A single measurement will yield a random bit.\\

To consider an analog of the twice watched pot, we may imagine two successive measurements on a single circulating qubit.
Consider the probability of finding the system in the state $|+\rr$ at one time, and $|\phi\rr$, at some earlier time.
First projecting the state $|\phi\rr$ onto $|+\rr$ gives a factor of $|\alpha^2|$. Just after a measurement result of $|+\rr$, the qubit is 'recycled' into the mixed state 
\bq
\rho_i = |\alpha|^2A(|+\rr\langle+|) = |\alpha|^2(1-k)|+\rr\langle +| + |\alpha|^2k|-\rr\langle -|
\eq
with a weight of
\bq
\omega(\phi) = \langle\phi|\rho_i|\phi\rr = |\alpha|^4(1-k) + |\alpha^2\beta^2|k
\eq
where we have described the noise $A$ as a linear operator on density matrices with an effective bit error rate $k$. Clearly it is proportional to the ordinary transition probability for a state $\phi$ to scatter onto $|+\rr$ and then back onto $\phi$. Here it is easy to see that integrating out $\phi$ will give the same weight for both classical measurement results. The result of the classical measurement is thus a random bit. Furthermore, we can see that the relative conditional probability of measuring the initial state $|\phi\rr$, given a later or earlier measurement of $|+\rr$ is the weight given here.

 \begin{figure}[h!]
  \caption{Transition probabilities for successive measurements and rotations of $\phi$ depend on future gates }
  \centering
    \includegraphics[width=0.4\textwidth]{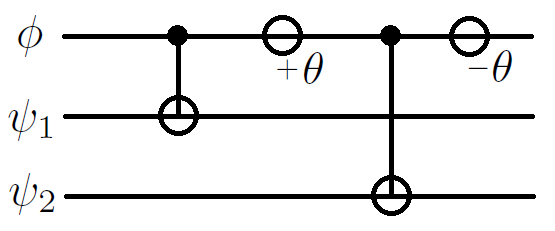}
\end{figure}

In the limit of small noise, two successive measurements of a qubit at a mutual angle $\theta$ will give the same result with probability 
\bq
\tilde{p} = \frac{p^2}{p^2 + (1-p)^2} = ((\tan{\theta})^4 +1)^{-1}.
\eq 
Notice the quartic term, as opposed to $\tan^2\theta$ that would appear without the CTC. This is a strong deviation from normal unitary evolution.
 We can model this scenario using a series of five gates acting on a single periodic channel and three probe channels.
Two rotation gates in between three $C_{not}$ measurement gates. Each measurement gate is controlled by the periodic bit and acts on a separate probe bit.
We will take each probe to be pre-selected in the $|0\rr$ state to isolate this effect from back-propagation. This gives us a loop out state of
\bq
|\psi_{out} = |\phi_{TM}\rr \otimes |000\rr
\eq
Acting with the first measurement we get
\bq
|\psi_{out}'\rr = U_{cn23}|\psi_{out}\rr = \frac1{\sqrt{2}}(|00000\rr + |11100\rr)
\eq
Next comes rotation by $\theta_1$.
\bq
|\psi_{out}''\rr = U_{\theta_1}|\psi_{out}'\rr = \frac{1}{\sqrt{2}}(\cos\theta_1|00000\rr + \cos\theta_1|11100\rr + \sin\theta_1|01000\rr - \sin\theta_1|10100\rr)
\eq
The second measurement maps the state to
\bq
|\psi_{out}'''\rr = U_{cn24}|\psi_{out}''\rr = \frac{1}{\sqrt{2}}(\cos\theta_1|00000\rr + \cos\theta_1|11110\rr + \sin\theta_1|01010\rr - \sin\theta_1|10100\rr)
\eq
The second rotation by angle $\theta_2$ gives
\bqn
|\psi_{out}''''\rr = U_{\theta_2}|\psi_{out}'''\rr\nb\\
 = \frac{1}{\sqrt{2}}(\cos\theta_1\cos\theta_2|00000\rr + \cos\theta_1\cos\theta_2|11110\rr + \sin\theta_1\cos\theta_2|01010\rr - \sin\theta_1\cos\theta_2|10100\rr)\nb\\
 +\cos\theta_1\sin\theta_2|01000\rr - \cos\theta_1\sin\theta_2|10110\rr - \sin\theta_1\sin\theta_2|00010\rr - \sin\theta_1\sin\theta_2|11100\rr) &
\eqn
Finally the in state is found after the third measurement
\bqn
|\psi_{in}\rr = U_{cn25}|\psi_{out}''''\rr\nb\\
 = \frac{1}{\sqrt{2}}(\cos\theta_1\cos\theta_2|00000\rr + \cos\theta_1\cos\theta_2|11111\rr + \sin\theta_1\cos\theta_2|01011\rr - \sin\theta_1\cos\theta_2|10100\rr)\nb\\
 +\cos\theta_1\sin\theta_2|01001\rr - \cos\theta_1\sin\theta_2|10110\rr - \sin\theta_1\sin\theta_2|00010\rr - \sin\theta_1\sin\theta_2|11101\rr) &
\eqn
The Bell projection of this state gives
\bq
|\bar{\psi}_B\rr = \lr \phi_{Bell} | \psi_{in} \rr = \frac12( \cos\theta_1\cos\theta_2|000\rr + \cos\theta_1\cos\theta_2|111\rr  - \sin\theta_1\sin\theta_2|010\rr - \sin\theta_1\sin\theta_2|101\rr) 
\eq
and normalization factor
\bq
N^2 = \frac12(\cos^2\theta_1\cos^2\theta_2 +\sin^2\theta_1\sin^2\theta_2)
\eq
Two things are of interest in this result. The first and third measurements are perfectly entangled. This is not normal behavior, but not completely unsurprising 
since from the looping channel's perspective measurements 1 and 3 occur back to back with no rotation in between. 
The stubborn spin effect manifests when we look at the probability of the intermediate transition. 
Just as in the spin case, the renormalized projected state vector gives a probability of observing alternating bits as 
\bq
p_{flip} = \frac{\sin^2\theta_1\sin^2\theta_2}{N^2} = \frac1{\cot^2\theta_1\cot^2\theta_2 +1}
\eq
This holds even without the third probe channel.
In the noisy case the three orthogonal projections and norms are
\bqn
|\phi_-\rr = \frac1{\sqrt{2}}( |00\rr - |11\rr)\\
|\phi_N\rr = \frac1{\sqrt{2}}(|01\rr + |10\rr)\\
|\phi_{-N} = \frac1{\sqrt{2}}(|01\rr - |10\rr)\\
|\bar{\psi}_-\rr = \lr \phi_- | \psi_{in} \rr = \frac12( \cos\theta_1\cos\theta_2|000\rr - \cos\theta_1\cos\theta_2|111\rr  - \sin\theta_1\sin\theta_2|010\rr + \sin\theta_1\sin\theta_2|101\rr) \\
|\bar{\psi}_N\rr = \lr \phi_N | \psi_{in} \rr = \frac12( \cos\theta_1\sin\theta_2|001\rr + \sin\theta_1\cos\theta_2|011\rr  - \sin\theta_1\cos\theta_2|100\rr - \cos\theta_1\sin\theta_2|110\rr) \\
|\bar{\psi}_{-N}\rr = \lr \phi_{-N} | \psi_{in} \rr = \frac12( \cos\theta_1\sin\theta_2|001\rr + \sin\theta_1\cos\theta_2|011\rr  + \sin\theta_1\cos\theta_2|100\rr + \cos\theta_1\sin\theta_2|110\rr) \\
N_-^2 = N^2\\
N_N^2 = N^2_{-N}= \frac12(\sin^2\theta_1\cos^2\theta_2 + \cos^2\theta_1\sin^2\theta_2)
\eqn
The partition function and density matrix are given by
\bqn
Z_\lambda= (1-\lambda)N^2 +\lambda/4\\
\rho = \frac{4 -3\lambda}{4Z_\lambda}|\bar{\psi}_B\rr\lr \bar{\psi}_B| + \frac\lambda{4Z_\lambda}\left( |\bar{\psi}_-\rr\lr \bar{\psi}_-| +|\bar{\psi}_N\rr\lr \bar{\psi}_N| + |\bar{\psi}_{-N}\rr\lr \bar{\psi}_{-N}| \right) 
\eqn
This results in a flip probability for the second measurement of 
\bq
P_{flip-\lambda} = \frac{\sin^2\theta_1}{2Z_\lambda}\left( (1-\lambda)\sin^2\theta_2 + \frac\lambda{2}\right)
\eq
The probability still depends nontrivially on the angle of the gate yet to be entered by the $\phi$ channel, but this effect decreases with noise. 
The classical limit gives a similar result.
 We use the leftmost bit as the classical periodic channel bit, and the middle and right bits as the first and second probes respectively.
The classical model in state is a mixture of the two possible unitary evolutions.
\bqn
U_t|000\rr = \cos\theta_1\cos\theta_2|000\rr + \sin\theta_1\cos\theta_2|101\rr + \cos\theta_1\sin\theta_2|100\rr - \sin\theta_1\sin\theta_2|001\rr\\
U_t|100\rr = \cos\theta_1\cos\theta_2|111\rr - \cos\theta_1\sin\theta_2|011\rr - \sin\theta_1\cos\theta_2|010\rr - \sin\theta_1\sin\theta_2|110\rr
\eqn
The classical weights are the norms of the projections of these states onto each classical state. The four decoherent projections are
\bqn
|\bar{\psi}_{00}\rr = \lr 0,xx| U_t | 000\rr =  \cos\theta_1\cos\theta_2|00\rr - \sin\theta_1\sin\theta_2|01\rr\\
|\bar{\psi}_{11}\rr = \lr 1,xx|U_t|100\rr =  \cos\theta_1\cos\theta_2|11\rr -\sin\theta_1\sin\theta_2|10\rr\\
|\bar{\psi}_{10}\rr = \lr 1,xx|U_t|000\rr = \sin\theta_1\cos\theta_2|01\rr + \cos\theta_1\sin\theta_2|00\rr \\
|\bar{\psi}_{01}\rr =\lr 0,xx|U_t|100\rr = - \cos\theta_1\sin\theta_2|11\rr - \sin\theta_1\cos\theta_2|10\rr
\eqn
The corresponding weights are
\bqn
\omega_{00} =\lr \bar{\psi}_{00} |\bar{\psi}_{00}\rr =  \omega_{11} =\lr \bar{\psi}_{11} |\bar{\psi}_{11}\rr =\cos^2\theta_1\cos^2\theta_2 +\sin^2\theta_1\sin^2\theta_2\\
\omega_{01} =\lr \bar{\psi}_{01} |\bar{\psi}_{01}\rr =  \omega_{10} =\lr \bar{\psi}_{10} |\bar{\psi}_{10}\rr =\sin^2\theta_1\cos^2\theta_2 + \cos^2\theta_1\sin^2\theta_2
\eqn
The partition function is defined by
\bq
Z = (1-k)(\omega_{00}+\omega_{11}) + k(\omega_{01}+\omega_{10}) 
\eq
and density matrix by
\bq
\rho_{cl} =  \frac{1-k}{Z_{2}}|\bar{\psi}_{00}\rr\lr\bar{\psi}_{00}| + \frac{k}{Z_{cl}}|\bar{\psi}_{01}\rr\lr \bar{\psi}_{01}| + \frac{1-k}{Z_{2}}|\bar{\psi}_{11}\rr\lr\bar{\psi}_{11}| + \frac{k}{Z_{cl}}|\bar{\psi}_{10}\rr\lr \bar{\psi}_{10}|
\eq
The flip probability for the classical limit channel is therefore
\bqn
P_{flip-cl} = \frac{(1-k)\sin^2\theta_1\sin^2\theta_2 + k\sin^2\theta_1\cos^2\theta_2}{\cos^2\theta_1\cos^2\theta_2 +\sin^2\theta_1\sin^2\theta_2}\nb\\
= (1-k)(1+ \cot^2\theta_1\cot^2\theta_2)^{-1} + k(\cot^2\theta_1+\tan^2\theta_2)^{-1}
\eqn
This decreased probability for detecting a change in the state of the post-selected system is fairly generic. A typical state will not be both an eigenstate of the measurement and of the circuit as a whole. As a result, multiple measurements often decrease the partition function of the system by reducing the normalization factor $N$. Consider the dragging of a spin by a large number of nearly parallel measurements, an example an anti Zeno effect. A grandfather circuit could be made if the dragging moves the state onto one that is orthogonal to the initially observed state of the channel. The spin will appear to be coupled to an unknown field that resists the dragging effect. It can also serve as the disentangling mechanism thought to be required in the final state model\cite{presk}\\


\section{Amnesia paradox}

This is essentially the time reverse of the unproven theorem paradox, in which two different input external states map onto the same external out state. The nonunitary map generated violates the second law by reducing the entropy of a maximally mixed qubit. Here we begin with an arbitrary qubit coming in on state $\psi_1$, which then acts as the control qubit in a controlled not operation onto the unknown state $\phi_{TM}$. Then the bits are swapped, sending our original $\psi_1$ into the loop to be projected onto $\phi_{TM}$. The external bit has now been XORed with the temporary copy of itself.  As usual the channels are from left to right, the reference qubit, the periodic channel $\phi$, and then any external qubits $\psi_i$.
\bq
|\psi_{out}\rr = |\phi_{Bell}\rr \otimes ( \alpha |0\rr + \beta |1\rr )
\eq
The unitary evolution is
\bqn
|\psi_{in}\rr = U_{cn23}U_{swap23}|\psi_{out}\rr \nb\\
= \frac1{\sqrt{2}}\left( \alpha|000\rr + \alpha|101\rr + \beta|011\rr + \beta|110\rr \right)
\eqn
The Bell projection of this state gives us
\bq
|\bar{\psi}_B\rr = \frac{\alpha+\beta}2|0\rr = N_B|0\rr
\eq
\begin{figure}[h!]
  \caption{An amnesia type circuit made from two controlled not gates.}
  \centering
    \includegraphics[width=0.4\textwidth]{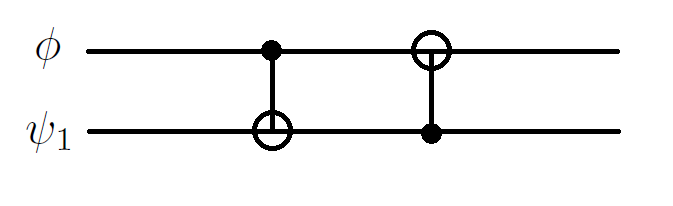}
\end{figure}
The scenario here is very similar to the measurement catastrophe. Destructive interference in the bit to be erased can make the normalization factor vanish.
The other three projections are
\bqn
|\bar{\psi}_-\rr = \frac{\alpha-\beta}2|0\rr =N_-|0\rr\\
|\bar{\psi}_N\rr = \frac{\alpha+\beta}2|1\rr =N_N|1\rr\\
|\bar{\psi}_{-N}\rr = \frac{\beta-\alpha}2|1\rr =N_{-N}|1\rr
\eqn
The mixture of these gives us the noisy channel approximation.
\bqn
Z_\lambda = (1-\lambda)N_B^2 + \lambda/4 =\frac14\left( \lambda + (1-\lambda)|\alpha+\beta|^2\right)\\
\rho_\lambda(\alpha) = \frac{1-\lambda}{4Z_\lambda}|\alpha+\beta|^2|0\rr\lr 0| + \frac{\lambda}{8Z_\lambda}I
\eqn
Just as with the measurement catastrophe, we can approach the singular point in the noisy approximation. 
In that case the result is a maximally mixed state for singular input.
In the classical limit we have the simple result.
\bqn
|\bar{\psi}_{00}\rr = \lr 0,x|U_{cn12}U_{swap12}|0,\psi_1\rr = \alpha|0\rr\\
|\bar{\psi}_{10}\rr = \lr 1,x|U_{cn12}U_{swap12}|0,\psi_1\rr = \beta|1\rr\\
|\bar{\psi}_{11}\rr = \lr 1,x|U_{cn12}U_{swap12}|1,\psi_1\rr = \beta|0\rr\\
|\bar{\psi}_{01}\rr = \lr 0,x|U_{cn12}U_{swap12}|1,\psi_1\rr = \alpha|1\rr\\
\omega_{00}=\omega_{01} = \alpha^2\\
\omega_{11}=\omega_{10} = \beta\beta^\dagger \\
Z_{cl} = (1-k)(\omega_{00}+\omega_{11}) + k(\omega_{01}+\omega_{10}) = 1\\
\rho_{cl} = (1-k)|0\rr\lr 0| + k|1\rr\lr 1|
\eqn
Where it appears that decoherence has eliminated the back-propagation effects, since $Z_{cl}$ is a constant.

This same process can be used to effectively erase entanglement with only local operations, another violation of unitarity.
 Adding a second external channel $\psi_2$ that begins maximally entangled with $\psi_1$, we have
\bq
|\psi_{out}\rr = |\phi_{Bell}\rr \otimes ( \alpha|00\rr + \beta|11\rr )
\eq
We can think of the bit to be erased as entangled with a distant channel $\psi_2$, so that none of the gates act on that channel.
The state after evolution is 
\bqn
|\psi_{in} \rr=  U_{cn23}U_{swap23}|\psi_{out}\rr \nb\\
= \frac1{\sqrt{2}}\left( \alpha|0000\rr + \alpha|1010\rr + \beta|0111\rr + \beta|1101\rr \right)
\eqn
The projected states are
\bqn
|\bar{\psi}_B\rr = \alpha|00\rr + \beta|01\rr =\frac12 |0\rr \otimes (\alpha|0\rr +\beta|1\rr)\\
|\bar{\psi}_-\rr = \alpha|00\rr - \beta|01\rr = \frac12|0\rr \otimes (\alpha|0\rr -\beta|1\rr)\\
|\bar{\psi}_N\rr =\alpha|10\rr + \beta|11\rr = \frac12 |1\rr \otimes (\alpha|0\rr +\beta|1\rr)\\\
|\bar{\psi}_{-N}\rr = -\alpha|10\rr + \beta|11\rr = \frac12|1\rr \otimes (-\alpha|0\rr +\beta|1\rr)\\
\eqn
All four of these states are product states. The breaking of entanglement creates a superposition of local states for the distant entangled channel $\psi_2$. \\

\begin{figure}[h!]
  \caption{Amnesia circuits acting on one of an entangled pair breaks the entanglement, resulting in a product of two pure states.}
  \centering
    \includegraphics[width=0.4\textwidth]{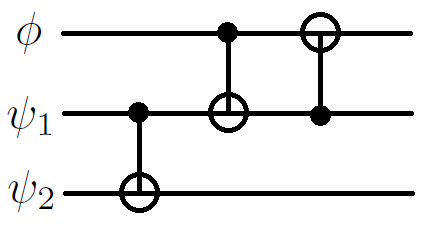}
\end{figure}
This can potentially be detected locally by an observer with access only to the $\psi_2$ channel.
 Since $\psi_2$ has no interactions in this circuit, the effects of the erasure of $\psi_1$ will back-propagate along $\psi_2$ until it interacts with something that dilutes or cancels the effect.
Just like the partition function for the classical version of the unentangled amnesia circuit, the normalization factor is constant.
\bq
N^2 = \frac14\alpha^2 + \frac14\beta\beta^\dagger = \frac14
\eq
This indicates no back-propagation onto the $\psi_1$ channel. 
The partition function is also constant
\bq
Z = \frac14(1-\lambda) + \frac\lambda{4} = \frac14\\
\eq
The density matrix has six nonzero components given by
\bq
\rho = \left( \begin{array}{cccc} (1-\frac\lambda{2})\alpha^2 & (1-\lambda)\alpha\beta & 0 & 0 \\\\  (1-\lambda)\alpha\beta^\dagger & (1-\frac\lambda{2})\beta\beta^\dagger & 0 & 0 \\\\ 0 & 0 & \frac\lambda{2}\alpha^2 &  0 \\\\ 0 & 0 & 0 & \frac\lambda{2}\beta\beta^\dagger \end{array} \right)
\eq
In addition to decoupling the two channels, the distant channel is slightly decohered in proportion to the noise of the erasing circuit. 
This is an important effect since it introduces noise into any communication between the erasing circuit and the distant channel $\psi_2$.
Such channels may be created by acting with a controlled phase flip on the $\psi_1$ channel before erasing it. 
This changes the sign of $\beta$ in the superposition of the $\psi_2$ channel.
Since the effect propagates back to the last interaction of the $\psi_2$ channel, this allows communication from the vicinity of the time loop to an arbitrary point in spacetime,
 such as the causal past of the creation of the loop itself, or across an event horizon.
 This could be regarded as a principle of non-censorship, although other effects may interfere with such signaling.\\

\begin{figure}[h!]
  \caption{The phase information of the broken entanglement becomes localized, and detectable with local measurements on the second channel. Since this phase can be affected by a controlled phase flip on the first channel, it leads to a secondary loop.}
  \centering
    \includegraphics[width=0.4\textwidth]{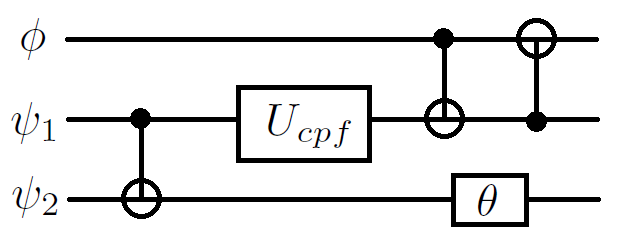}
\end{figure}

The classical version of the circuit with entangled external channels is given by
\bqn
|\psi_{ent}\rr = \alpha|00\rr + \beta|11\rr\\
|\bar{\psi}_{00}\rr = \lr 0,x|U_{cn12}U_{swap12}|0,\psi_{ent}\rr = \alpha|00\rr\\
|\bar{\psi}_{10}\rr = \lr 1,x|U_{cn12}U_{swap12}|0,\psi_{ent}\rr = \beta|11\rr\\
|\bar{\psi}_{11}\rr = \lr 1,x|U_{cn12}U_{swap12}|1,\psi_{ent}\rr = \beta|01\rr\\
|\bar{\psi}_{01}\rr = \lr 0,x|U_{cn12}U_{swap12}|1,\psi_{ent}\rr = \alpha|10\rr\\
\omega_{00}=\omega_{01} = \alpha^2\\
\omega_{11}=\omega_{10} = \beta\beta^\dagger \\
Z_{cl} = (1-k)(\omega_{00}+\omega_{11}) + k(\omega_{01}+\omega_{10}) = 1\\
\rho_{cl} = (1-k)\left( \alpha^2|00\rr\lr 00| + \beta\beta^\dagger|01\rr\lr 01| \right) + k\left( \alpha^2|10\rr\lr 10| + \beta\beta^\dagger|11\rr\lr 11|\right)
\eqn
In the classical case the distant channel is completely decohered along the preferred basis states, but its eigenvalues remain the same, so no change will be detected locally.
Likewise the addition of a controlled phase flip acting on $\psi_1$ before the swap with $\psi$ fails to communicate information to $\psi_2$.\\

Returning to the case of measurement catastrophe in the unproven proof circuit, it was noted that this is an amnesia type circuit. 
The same decoupling behavior for entangled incoming probe states is also present in the single $C_{not}$ gate circuit. Let
\bq
|\psi_{out}\rr = |\phi_{Bell}\rr \otimes ( \alpha|00\rr + \beta|11\rr )
\eq
and 
\bqn
|\psi_{in}\rr = U_{cn23}|\psi_{out}\rr\nb\\
= \frac1{\sqrt{2}}\left( \alpha |0000\rr + \alpha|1110\rr + \beta|0011\rr + \beta|1101\rr \right)
\eqn
Two projections are nonzero.
\bqn
|\bar{\psi}_B\rr = \frac{\alpha}{2}\left( |00\rr + |10\rr \right) + \frac{\beta}{2}\left( |11\rr + |01\rr\right) = \frac12\left( |0\rr + |1\rr\right) \otimes \left( \alpha|0\rr + \beta|1\rr \right)\\
|\bar{\psi}_-\rr  =  \frac{\alpha}{2}\left( |00\rr - |10\rr \right) + \frac{\beta}{2}\left( |11\rr - |01\rr\right) = \frac12\left( |0\rr - |1\rr\right) \otimes \left( \alpha|0\rr - \beta|1\rr \right)\\
N_B^2=N_-^2=1/4
\eqn
The standard mixture gives the noisy channel.
\bqn
Z_\lambda = (1-\lambda)N^2 + \lambda/4 = 1/4\\
\rho =(4 -3\lambda)|\bar{\psi}_B\rr\lr \bar{\psi}_B| + \lambda  |\bar{\psi}_-\rr\lr \bar{\psi}_-|
\eqn
Just as in the other case, the noise propagate to the decoupled state as decoherence.\\

The secondary loop circuit consists of a Bell pair, $\pi/4$ rotation on one channel and a conditional phase flip on the other. The second channel then goes into an amnesia circuit creating a product state. If the control of the conditional phase flip begins as an arbitrary product state we have
\bq
|\psi_{out}\rr = |\phi_{TM}\rr \otimes \frac1{\sqrt{2}}\left( |00\rr + |11\rr \right) \otimes \left( \alpha|0\rr + \beta|1\rr \right)
\eq

The secondary channel is created by the action of four gates

\bq
|\psi_{in}\rr = U_{cn32}U_{cn23}U_{cfp53}U_{rot4}|\psi_{out}\rr
\eq
The intermediate states are
\bqn
U_{rot4}|\psi_{out}\rr = |\phi_{TM}\rr \otimes \frac12\left( |00\rr - |01\rr +|10\rr + |11\rr\right) \otimes \left( \alpha|0\rr + \beta|1\rr\right)\\
U_{cfp53}U_{rot4}|\psi_{out}\rr = \nb\\
 |\phi_{TM}\rr \otimes \frac12 \left( \alpha|000\rr -\alpha|010\rr + \alpha|100\rr + \alpha|110\rr +\beta|001\rr -\beta|011\rr -\beta|101\rr - \beta|111\rr \right)\\
\sqrt{8}|\psi_{in}\rr = \alpha|00000\rr -\alpha|00010\rr + \alpha|01100\rr + \alpha|01110\rr \nb\\
+\beta|00001\rr -\beta|00011\rr -\beta|01101\rr - \beta|01111\rr \nb\\
\alpha|10100\rr -\alpha|10110\rr + \alpha|11000\rr + \alpha|11010\rr \nb\\
+\beta|10101\rr -\beta|10111\rr -\beta|11001\rr - \beta|11011\rr
\eqn
and the projected states are
\bqn
\bar\psi_B = \frac14\left(\alpha|000\rr  -\alpha|010\rr +\beta|001\rr -\beta|011\rr\right)\nb\\
+\frac14\left(\alpha|000\rr + \alpha|010\rr - \beta|001\rr - \beta|011\rr\right)\nb\\
= \frac12\left( \alpha|000\rr -\beta|011\rr \right)\\ 
\bar\psi_N = \frac12\left( \alpha|100\rr -\beta|111\rr \right)\\ 
\bar\psi_- =  \frac14\left(\alpha|000\rr  -\alpha|010\rr +\beta|001\rr -\beta|011\rr\right)\nb\\
-\frac14\left(\alpha|000\rr + \alpha|010\rr - \beta|001\rr - \beta|011\rr\right)\nb\\
= \frac12\left( \beta|001\rr -\alpha|010\rr \right)\\
\bar\psi_{-N}= \frac12\left( \beta|101\rr -\alpha|110\rr \right)
\eqn

The movement of the relative amplitudes $\alpha$ and $\beta$ from the $\psi_3$ channel to the entangled state with the $\psi_2$ channel, without direct interaction indicates that classical measurements of $\psi_2$ obtain information about $\psi_3$, up to the limit of the rate of phase errors in the $\phi$ channel. The mixed state is

\bqn
Z = \frac\lambda{4} + \left(1-\lambda\right)N_B^2 = \frac14\\
\rho = Z^{-1}\left( 1-\frac34\lambda\right)|\bar\psi_B\rr\lr\bar\psi_B| + \frac\lambda{4Z}|\bar\psi_-\rr\lr\bar\psi_-| + \frac\lambda{4Z}|\bar\psi_N\rr\lr\bar\psi_N| + \frac\lambda{4Z}|\bar\psi_{-N}\rr\lr\bar\psi_{-N}| 
\eqn

The 'second order' grandfather paradox can be formed by carefully choosing the entangled states. If we take the measurement of $\psi_2$ as the control of the phase flip of the $\psi_1$ channel, we create the particular two channel state

\bqn
|\psi_{GF}\rr = U_{cpf21}U_{rot2}|\psi_{Bell}\rr\nb\\
= \frac12 \left( |00\rr + |01\rr - |10\rr - |11\rr \right) \nb\\
= \frac12 \left( |0\rr - |1\rr \right) \otimes \left( |0\rr + |1\rr \right)
\eqn

The grandfather paradox of the secondary channel is really just the creation of a measurement catastrophe in the amnesia circuit. The creation of secondary channels and local interference effects of a final state projection black hole has been investigated also by \cite{yuri1,yuri2}. Because these secondary channels are not limited to the forward light cone, they can be used to communicate with observers outside the black hole.


\section{Back-propagation}

In an ordinary experiment, one of the first things to understand is the idea of controlled conditions. 
The reproducibility of results, the assumption of the ability to do independent tests, and the concept of a prepared initial state of the system
 all stem in some way from our basic intuitions of causality. A system that does not respect these can reasonably be called acausal.
It would be tempting to regulate the effects of projection to a special spacelike surface such as an event horizon, but this would also be misleading.
One reason is because of the way that entanglement acts with projection. States that are highly entangled with a projected channel to the point 
that we might begin considering them as classical states, even then can be projected into very different mixtures.
 Consider the controlled not grandfather circuit. When we prepare the control channel in an eigenstate, we are selecting the initial value
 in much the same way as when we choose to accept in the ensemble only runs that satisfy the post-selection measurement. 
If the two measurements are not consistent with each other, then the ensemble is empty.
 The size of the ensemble is proportional to the normalization factor. 
It can be thought of as a measure of mutual consistency between preselected and post-selected states. 
One could imagine that the projection surface acts on the control channel and forces it into the state $|0\rr$, but it must also force any
 states entangled with that channel as well. The extent of this is such that we would not be able to point to the surface at which the
 projection occured, since any record would also be projected.
 The difference between such extensive reshaping of the wave-function and true backwards causal action may be impossible to detect even in principle.
Let us return to the controlled not circuit, but add a few additional probe states. 
We can use these to monitor the value of the control qubit before and after it passes through the $C_{not}$ and acts on the peridic qubit $\phi$.
The product state is
\bqn
|\psi_{out}\rr = |\phi_{Bell}\rr \otimes \frac1{\sqrt{2}}\left( | 0\rr + |1\rr \right) \otimes |0\rr \otimes |0\rr \nb\\
= \frac12 \left( |00000\rr + |11000\rr + |00100\rr + |11100\rr\right)
\eqn
Here we have put the control channel in the Hadamard state, and the last two channels are probes in the $|0\rr$ state.
The action of the circuit will sandwich the 'gun' $C_{not}$ between two measurement gates.
\bqn
|\psi_{in}\rr = U_{cn35}U_{cn32}U_{cn34}|\psi_{out}\rr \nb\\
= \frac12\left( |00000\rr + |11000\rr + |01111\rr + |10111\rr \right)
\eqn
Then projections give us 
\bqn
|\bar{\psi}_B\rr = \frac1{\sqrt{2}}|000\rr \\
|\bar\psi_N\rr = \frac1{\sqrt{2}}|111\rr \\
|\bar\psi_-\rr=|\bar\psi_{-N}\rr = 0\\
N^2 = \frac12
\eqn
This clearly shows that the value of the control bit, as far as entangled observers are concerned does not change in the proximity of the the circuit.
Further, that this holds either with or without noise, as well as in a classical limit for the periodic bit. 
By specifying the initial state of the probe and control qubits, we manually blocking the effects of post-selection from propagating further back. 
The the amplitudes of states in post-selection quantum mechanics back-propagate like the probabilities of a Bayesian model. 
If we follow the control channel further back, we can observe behavior similar to that of the periodic channel itself. 
A grandfather circuit with a controlled not acting on an input channel behaves much like a simple projection of that channel onto the $|0\rr$ state would.
Consider a stubborn spin circuit combined with a controlled not paradox circuit. 
The system will contain the periodic channel $\phi$, the control channel and a probe channel.
Before the emergence of the $\phi$ state from it's loop, we take a the control and probe channels to be in the state
\bq
|\psi_{34}\rr = |00\rr
\eq
To simulate a measurement at an angle $\theta$, we apply a $C_{not}$ sandwiched between two rotations, one by $\theta$ and one by $-\theta$.
\bqn
|\psi_{34}'\rr = U_{rot3}(-\theta)U_{cn34}U_{rot3}(\theta)|\psi_{34}\rr \nb\\
= \cos^2\theta|00\rr + \sin\theta\cos\theta|11\rr - \sin\theta\cos\theta|10\rr + \sin^2\theta|01\rr
\eqn
The controlled grandfather circuit then effectively projects the control qubit against $\lr0|$, giving
\bq
|\bar\psi_B\rr = \cos^2\theta|00\rr + \sin^2\theta|01\rr
\eq
If the original circuit contained noise, then the equivalent projection onto the control channel input will contain at least as much noise. 
The skewing effect of post-selection need not be as extreme as we follow the system backwards.
 Noise acting on the control channel before it acts on the periodic channel can have a screening effect, as can certain configurations of gates.
If we change this setup so that the grandfather circuit is replaced by a faulty gun, then the control channel will only inherit some of the selective bias.
We replace the final controlled not gate from the control bit into the periodic channel with a controlled rotation by an angle $\theta_g$.
 The angle of the stubborn spin gates shall be $\theta_s$.
The product state is
\bq
|\psi_{out}\rr = |\phi_{Bell}\rr \otimes |00\rr
\eq
After acting with the first three gates this gives
\bqn
|\psi'\rr =  U_{rot3}(-\theta_s)U_{cn34}U_{rot3}(\theta_s)|\psi_{out}\rr \nb\\
= |\phi_{Bell}\rr \otimes \left ( \cos^2\theta_s|00\rr + \sin\theta_s\cos\theta_s|11\rr - \sin\theta_s\cos\theta_s|10\rr + \sin^2\theta_s|01\rr \right)
\eqn
Then the action of the faulty gun circuit gives
\bqn
|\psi_{in}\rr = U_{crot32}(\theta_g)|\psi'\rr\\
=|\phi_{Bell}\rr \otimes \left ( \cos^2\theta_s|00\rr + \sin^2\theta_s|01\rr \right) \nb\\
+ \left( \cos\theta_g|\phi_{Bell}\rr + \frac{\sin\theta_g}{\sqrt{2}}( |01\rr - |10\rr ) \right) \otimes \left(\sin\theta_s\cos\theta_s|11\rr - \sin\theta_s\cos\theta_s|10\rr \right)
\eqn
Now the projections are
\bqn
|\bar\psi_B\rr = \cos^2\theta_s|00\rr + \sin^2\theta_s|01\rr + \cos\theta_g\sin\theta_s\cos\theta_s|11\rr - \cos\theta_g\sin\theta_s\cos\theta_s|10\rr\\
|\bar\psi_{-N}\rr = \sin\theta_g\sin\theta_s\cos\theta_s|11\rr - \sin\theta_g\sin\theta_s\cos\theta_s|10\rr \\
|\bar\psi_N\rr=|\bar\psi_-\rr = 0
\eqn
and the normalization factor is
\bq
N_B^2= 1-\sin^2\theta_g\sin^2\theta_s\cos^2\theta_s
\eq
This means that without noise we have a transition probability for the control qubit of
\bq
P_{flip} = \frac1{N^2}|\lr 01 | \bar\psi_B\rr |^2 + \frac1{N^2}|\lr 11 | \bar\psi_B\rr |^2 = \sin^2\theta_s \left(\frac{1-\sin^2\theta_g\cos^2\theta_s}{1-2\sin^2\theta_g\sin^2\theta_s\cos^2\theta_s}\right)
\eq
When $\theta_g=0$ the control qubit has no interaction with $\phi$, and normal behavior is restored. At $\theta_g=\pi/2$ the circuit behaves as though the projection were acting on the control bit as before. 
The angle $\theta_g$ controls to some degree how much the projection back-propagates onto the control channel. This is incidentally also how much influence the control channel has over $\phi$.
If we remove the control by another step, so that it acts with $U_{crot}(\theta)$ on a second control channel which then acts with another similar gate on the periodic channel, we can see that some of the
stubborn spin force is dissipated into the intermediate channels. The full calculation is
\bqn
|\psi_{out}\rr = |\phi_{Bell}\rr \otimes |000\rr\\
|\psi'\rr =  U_{rot4}(-\theta_s)U_{cn45}U_{rot4}(\theta_s)|\psi_{out}\rr \nb\\
= |\phi_{Bell}\rr \otimes \left ( \cos^2\theta_s|000\rr + \sin\theta_s\cos\theta_s|011\rr - \sin\theta_s\cos\theta_s|010\rr + \sin^2\theta_s|001\rr \right)\\
|\psi_{in}\rr = U_{crot32}(\theta_{g2})U_{crot43}(\theta_{g1})|\psi'\rr\\
=|\phi_{Bell}\rr \otimes \left ( \cos^2\theta_s|000\rr + \sin^2\theta_s|001\rr \right) \nb\\
+ |\phi_{Bell}\rr \otimes \cos\theta_{g1}\sin\theta_s\cos\theta_s\left(|011\rr - |010\rr \right)\nb\\
+  |\phi_{Bell}\rr \otimes \cos\theta_{g2}\sin\theta_{g1}\sin\theta_s\cos\theta_s\left(|111\rr - |110\rr \right)\nb\\
+  \frac1{\sqrt{2}}\left( |01\rr - |10\rr \right) \otimes \sin\theta_{g1}\sin\theta_{g2}\sin\theta_s\cos\theta_s\left(|111\rr - |110\rr \right)
\eqn
Th right hand bit probes whether a the first control is flipped by the first rotation. The middle bit is the first control, which acts on the left bit which then acts on $\phi$.
The projection without noise eliminates only the last four terms.
\bqn
|\bar\psi_B\rr = \left ( \cos^2\theta_s|000\rr + \sin^2\theta_s|001\rr \right) \nb\\
+ \cos\theta_{g1}\sin\theta_s\cos\theta_s\left(|011\rr - |010\rr \right)\nb\\
+  \cos\theta_{g2}\sin\theta_{g1}\sin\theta_s\cos\theta_s\left(|111\rr - |110\rr \right)
\eqn
The normalization factor and stubborn spin flip probability are
\bqn
N^2 = 1 - 2\sin^2\theta_s\cos^2\theta_s\sin^2\theta_{g1}\sin^2\theta_{g2} \\
P_{flip} = \sin^2\theta_s\left( \frac{1  - \cos^2\theta_s\sin^2\theta_{g1}\sin^2\theta_{g2}}{1 - 2\sin^2\theta_s\cos^2\theta_s\sin^2\theta_{g1}\sin^2\theta_{g2}}\right)
\eqn
This result is equivalent to the substitution
\bq
\sin^2\theta_g \rightarrow \sin^2\theta_{g1}\sin^2\theta_{g2}
\eq
as far as deviation from normal behavior for the probe qubit is concerned. The more interactions between a given probe channel and the more diluted the effects of post-selection typically are.
For the two step control circuit the classical limit is
\bqn
\omega_{00} = \omega_{11} = N^2 \\
\omega_{01} = \omega_{10} = 2 \sin^2\theta_{g1}\sin^2\theta_{g2}\sin^2\theta_s\cos^2\theta_s
\eqn
The noisy classical and coherent partition functions are
\bqn
Z_{cl}=(1-k)(\omega_{00}+\omega_{11}) + k (\omega_{01} + \omega_{10} ) = 2(1-k)N^2 + 2k(1-N^2) \nb\\
= 2 - 2k -4(1-2k)\sin^2\theta_{g1}\sin^2\theta_{g2}\sin^2\theta_s\cos^2\theta_s\\
Z_\lambda = (1-\lambda)N^2 + \lambda/4\nb\\
= 1 - \frac34\lambda  -2(1-\lambda)\sin^2\theta_{g1}\sin^2\theta_{g2}\sin^2\theta_s\cos^2\theta_s
\eqn
To find the density matrix and flip probability in these noisy cases we need the complementary projection
\bqn
|\bar\psi_{-N}\rr = \sin\theta_{g1}\sin\theta_{g2}\sin\theta_s\cos\theta_s\left(|111\rr - |110\rr \right)\\
= \lr 1 ,xxx| U_t | 0 ,\psi \rr = \lr 0,xxx| U_t | 1,\psi \rr
\eqn
The density matrices are
\bqn
\rho_{cl} = \frac{1-k}{Z_{cl}}|\bar\psi_B\rr\lr \bar\psi_B | + \frac{k}{Z_{cl}}|\bar\psi_{-N}\rr\lr \bar\psi_{-N}|\\
\rho_\lambda =\frac{4-3\lambda}{4Z_\lambda}|\bar\psi_B\rr\lr \bar\psi_B | + \frac{\lambda}{4Z_\lambda}|\bar\psi_{-N}\rr\lr \bar\psi_{-N}|
\eqn
These give flip probabilities of
\bq
P_{cl-flip} = \sin^2\theta_s\left(\frac{1-k-(1-2k)\sin^2\theta_{g1}\sin^2\theta_{g2}\cos^2\theta_s}{1-k-2(1-2k)\sin^2\theta_{g1}\sin^2\theta_{g2}\sin^2\theta_s\cos^2\theta_s}\right)
\eq
and
\bq
P_{\lambda-flip} =  \sin^2\theta_s\left(\frac{4-3\lambda -(4-2\lambda)\sin^2\theta_{g1}\sin^2\theta_{g2}\cos^2\theta_s}{4-3\lambda-8(1-\lambda)\sin^2\theta_{g1}\sin^2\theta_{g2}\sin^2\theta_s\cos^2\theta_s}\right)
\eq

Another case to study is to add additional controlled not channels directly to the periodic channel. 
The flip probability will have a stubborn spin form when the sum of the other control channels is even, and a complementary behavior when the sum of the other channels is odd. 
When tracing over the other control channels the probability for an odd total acts like the noise parameter $k$ in classical limit. 
For the N-controlled not circuit we have
\bq
|\psi_{out}\rr = |\phi_{Bell}\rr \otimes |\psi_0\rr \otimes |\psi_1\rr \otimes ... \otimes |\psi_m\rr
\eq
After each other channel acts with $C_{not}$ on $\phi$,
\bqn
|\psi_{in}\rr = |\phi_{Bell}\rr \otimes \left( \alpha_0|0\rr \otimes |\psi_{even}\rr + \beta_0|1\rr \otimes |\psi_{odd}\rr \right) \nb\\
+ \frac{1}{\sqrt{2}}\left( |01\rr + |10\rr \right) \otimes \left( \alpha_0|0\rr \otimes |\psi_{odd}\rr + \beta_0|1\rr \otimes |\psi_{even}\rr\right)
\eqn
Where we have taken the even and odd subscripts to refer to the even and odd combinations of the other channels.
The projections are
\bqn
|\bar\psi_B\rr =  \alpha_0|0\rr \otimes |\psi_{even}\rr + \beta_0|1\rr \otimes |\psi_{odd}\rr \\
|\bar\psi_N\rr =  \alpha_0|0\rr \otimes |\psi_{odd}\rr + \beta_0|1\rr \otimes |\psi_{even}\rr \\
|\bar\psi_-\rr = |\bar\psi_{-N}\rr =0
\eqn
Tracing over the remaining channels we have
\bqn
E^2 = \lr \psi_{even} | \psi_{even} \rr \\
D^2 = \lr \psi_{odd} | \psi_{odd}\rr \\
N_B^2 = E^2\alpha_0^2 + D^2\beta_0\beta_0^\dagger \\
N_N^2 =  D^2\alpha_0^2 + E^2\beta_0\beta_0^\dagger 
\eqn
The closer $E^2$ gets to $1/2$, the closer $N_B^2$ becomes to constant. This indicates less back-propagation onto the $\psi_0$ channel for $E^2$ closer to $1/2$.
If we add another control channel, then the value of $E^2$ will in general become closer to $1/2$ for any new product state we choose.
\bqn
E_{m+1}^2 = E_m^2\alpha_{m+1}^2 + D_m^2(1-\alpha_{m+1}^2)\\
D_{m+1}^2 = D_m^2\alpha_{m+1}^2 + E_m^2(1-\alpha_{m+1}^2)
\eqn
Comparing this to the noisy classical controlled not circuit we find
\bq
k_{eff} = D^2
\eq
Furthermore we can see the strictly increasing effective noise by looking at the quantity $\epsilon$, which strictly decreases with extra channels.
\bqn
\epsilon_{m} = |E_m^2-\frac12| = (2\alpha_{m-1}^2 - 1)\epsilon_{m-1} \\
k_{eff} = \frac12 - \epsilon
\eqn
Since we can model noise in the control channel as the action of gates by hidden additional channels, 
we can see that noise in between our measurement of the stubborn spin on
 the control channel and the $C_{not}$ onto the periodic channel 
will partially shield the control channel at earlier times from stubborn spin like effects. 
Just as noise in a normal circuit decreases the effect of the initial state on post noise measurements, future noise
 shields present measurements from post-selection bias.
In general, if we expand a particular circuit by adding gates to the input channels, 
we can observe irregular behavior further and further prior to the circuit's interaction with the periodic qubits. 
In a sense, preparing a precise input state for a postselected circuit is no more trivial than guaranteeing a particular output value.
The degree to which these effects would otherwise back-propagate can be seen in the dependence of the partition function on the input channel states.
In the controlled grandfather circuit without noise, the partition function is equal to the normalization factor squared, and thus the square of the $|0\rr$ component of the control qubit.
In order to preselect the the control channel, we take a mixed qubit and perform a measurement to project it onto an eigenstate which we wish to use as the control input.
We could say the post-selection failed because our prepared control state was $|1\rr$, or the converse. 
In all cases where post-selection succeeded, we were unable to prepare the control state.
A measurement of our supposedly maximally mixed input state does not give the appropriate statistics. 
Instead it behaves as a weighted ensemble, where the population of each pure state input is given by the partition function of the circuit.
The resulting skewed density matrix is a measure of the input bias of the circuit. For completeness this is given as
\bq
\bar\rho = \frac{\int Z(\psi)|\psi\rr\lr\psi| d\psi}{\int Z d\psi}
\eq
This is of course the same form as for the density matrix of final outgoing values for the non-periodic channels.
 In a sense, both the initial and final values of the entangled channels are 'outputs' of the circuit, which is simply in a thermal state constrained by the pre and post-selected boundary conditions.
It should be noted that the stubborn spin and all the other pseudo-forces introduced by post-selection are $\emph{entropic forces}$. 


\section{Entropy renormalized}

In several cases we have seen that selection can reduce the entropy of a system. The treatment here is given in \cite{me,me2}, but is included again here for completeness.
 In the grandfather paradox unique states for the control qubits, and in the amnesia paradox, unique pure out states.
In order to describe larger systems and better understand the behavior post-selection induces it s important to quantify this effect.
If we limit ourselves to two qubit gates, the answer is simple. With four orthogonal projection states in the Bell model and two in the classical limit,
we can fix the value of two and one control qubits respectively.
 In the classical grandfather circuit a single $C_{not}$ will fix the value of its control qubit to $|0\rr$. 
In the coherent version of the circuit we can add a controlled phase flip which will accomplish the same for its control qubit.
The  $ C_{pf}+C_{not}$ circuit is described by
\bqn
|\psi_{out}\rr = |\phi_{Bell}\rr \otimes ( \alpha_1|0\rr + \beta_1|1\rr ) \otimes ( \alpha_2|0\rr + \beta_2|1\rr ) \\
|\psi_{in}\rr = U_{cpf21}U_{cnot31}|\psi_{out}\rr =\nb\\
\left( U_{not}|\phi\rr \otimes \alpha_1\beta_2|01\rr \right) +\left( U_{pf}|\phi\rr \otimes \beta_1\alpha_2|10\rr \right) + \left( U_{rot}|\phi\rr \otimes \beta_1\beta_2|11\rr \right) + \left(|\phi\rr \otimes \alpha_1\alpha_2|00\rr \right)
\eqn
Each $\phi$ part is one of the four orthogonal Bell states. Taking the projection will single out one of these four joint states for the control qubits.
 In the noiseless case only the $|00\rr$ term will survive.
Adding additional controlled not and controlled phase flips dilutes this fixing effect over multiple qubits.
This is shown for $C_{not}$ in the third party paradox, and for $C_{pf}$ in the null space of the Bell projection of the in state for the
twice watched pot circuits.
The situation changes with the introduction of the NAND, or Deutch gates. 
In these cases any number of external qubits can be fixed up to the level of noise.
Consider a circuit consisting of the periodic channel $\phi$ and two outside channels $\psi_1$, and $\psi_2$.
The double controlled rotation gate will take the $\psi_1$ and $\psi_2$ as controls and act on the state $\phi$ conditionally with the rotation $\theta_1$.
Then $\phi$ is further rotated by $\theta_2$ unconditionally.
The states are
\bqn
|\psi_{out}\rr = |\phi\rr \otimes ( \alpha_1|0\rr + \beta_1|1\rr ) \otimes ( \alpha_2|0\rr + \beta_2|1\rr ) \\
|\psi_{in}\rr = U_{r1}(\theta_2)U_{ccr231}(\theta_1)|\psi_{out}\rr =\nb\\
\left( U_r(\theta_2)|\phi\rr \right) \otimes \left(\alpha_1\alpha_2|00\rr + \alpha_1\beta_2|01\rr + \beta_1\alpha_2|10\rr \right) + \left(U_r(\theta_1+\theta_2)|\phi\rr\right) \otimes \beta_1\beta_2|11\rr
\eqn
For the Bell state model we have
\bqn
|\phi\rr = \frac1{\sqrt{2}}\left( |00\rr + |11\rr \right)\\
U_r(\theta)|\phi\rr = \frac1{\sqrt{2}}\left( \cos\theta|00\rr + \cos\theta|11\rr + \sin\theta|01\rr - \sin\theta|10\rr \right)
\eqn
The Bell projection of each part of the wave-function is proportional to the cosine of the total rotation of the $\phi$ part.
The four projections are,
\bqn
|\bar{\psi}_B\rr = \cos\theta_2 \left(\alpha_1\alpha_2|00\rr + \alpha_1\beta_2|01\rr + \beta_1\alpha_2|10\rr\right) + \cos(\theta_1+\theta_2)\beta_1\beta_2|11\rr\\
|\bar{\psi}_-\rr = |\bar{\psi}_{N}\rr = 0 \\
|\bar{\psi}_{-N}\rr = \sin\theta_2\left(\alpha_1\alpha_2|00\rr + \alpha_1\beta_2|01\rr + \beta_1\alpha_2|10\rr\right) + \sin(\theta_1+\theta_2)\beta_1\beta_2|11\rr\\
\eqn
By choosing the angles $\theta_1$ and $\theta_2$ we can select or deselect the $|11\rr$ state from the ensemble.
 Setting $\theta_1=\theta_2=\pi/2$, we have 
\bqn
|\bar{\psi}_B\rr = -\beta_1\beta_2|11\rr\\
|\bar{\psi}_{-N}\rr = \alpha_1\alpha_2|00\rr + \alpha_1\beta_2|01\rr + \beta_1\alpha_2|10\rr
\eqn
Then the noisy case is given by
\bqn
N^2 = \beta_1\beta_1^\dagger\beta_2\beta_2^\dagger\\
Z = (1-\lambda)N^2 + \lambda/4\\
\rho = (1-\frac34\lambda)\frac{N^2}Z|11\rr\lr 11| + \frac\lambda{4Z}|\bar{\psi}_{-N}\rr\lr \bar{\psi}_{-N}|
\eqn
The classical weights behave similarly,
\bqn
\omega_{00} = \omega_{11} = \beta_1\beta_1^\dagger\beta_2\beta_2^\dagger\\
\omega_{10} = \omega_{01} = 1 - \omega_{00}
\eqn
for the same choice of $\theta_1$ and $\theta_2$. The classical limit partition function is 
\bq
Z_{cl} = 2(1-k)N^2 + 2k(1-N^2)
\eq
and density the matrix
\bq
\rho_{cl} = \frac{2(1-k)N^2}Z |11\rr\lr 11| +   \frac{2k}Z|\bar{\psi}_{-N}\rr\lr \bar{\psi}_{-N}|
\eq
These results quickly extend to larger numbers of control qubits
In the case of a triple controlled rotation gate given by
\bqn
U_{cccr}(\theta) = I - (1-\cos\theta)|1111\rr\lr 1111| - (1-\cos\theta
)|1110\rr\lr 1110| \nb\\
+ \sin\theta|1111\rr \lr 1110| - \sin\theta|1110\rr \lr 1111|
\eqn
we can select the $|111\rr$ state, giving the projections,
\bqn
|\psi_{123}\rr =  ( \alpha_1|0\rr + \beta_1|1\rr ) \otimes ( \alpha_2|0\rr + \beta_2|1\rr ) \otimes ( \alpha_3|0\rr + \beta_3|1\rr )\\
|\bar{\psi}_B\rr = \beta_1\beta_2\beta_3|111\rr\\
|\bar{\psi}_-\rr = |\bar{\psi}_{N}\rr = 0 \\
|\bar{\psi}_{-N}\rr = |\psi_{123}\rr - |\bar{\psi}_B\rr
\eqn
and mixture of
\bqn
N^2 =  \beta_1\beta_2\beta_3 \beta_1^\dagger\beta_2^\dagger\beta_3^\dagger\\
Z = (1-\lambda)N^2 + \lambda/4 \\
\rho = \left(1-\frac34\lambda\right)\frac{N^2}Z |111\rr\lr 111| + \frac{\lambda}{4Z}|\bar{\psi}_{-N}\rr\lr \bar{\psi}_{-N}|
\eqn
For any number of control qubits, the mixed state gives us a skew factor for the probability. Defining the coefficients as
\bq
\rho = a|111\rr\lr 111| + b|\bar{\psi}_{-N}\rr\lr \bar{\psi}_{-N} | 
\eq
comparing the ratio for the pure product state $a/b$,  to the skewed ratio of the mixture of projected states ${\bar{a}}/{\bar{b}}$, gives
\bq
\Omega_\lambda = \frac{b}{a} \cdot \bar{a}/\bar{b} = \frac4\lambda -3
\eq
for the noisy Bell model, and
\bq
\Omega_k = \frac{b}{a} \cdot \bar{a}/\bar{b} = \frac1k -1
\eq
for the classical limit.
The entropy of a mixed ensemble with $n+1$ possible states can be written as
\bq
-S_0 = a \ln a + \sum_i^n b_i \ln b_i
\eq
The entropy of the skewed ensemble created by selecting the state associated with $a$ is
\bqn
-S_\Omega = \bar{a}\ln \bar{a} + \sum_i^n \bar{b}_i \ln \bar{b}_i \nb\\
= \frac{\Omega a}{Z'} \ln\left(\frac{\Omega a}{Z'}\right) + \sum_i^n\frac{b_i}{Z'} \ln\left(\frac{b_i}{Z'}\right)
\eqn
where we have defined
\bq
Z' = \left(\Omega - 1 \right) a +1
\eq
We are interested in the difference of the two entropies.
\bqn
\Delta S = \bar{S}-S_0 = \ln Z' - \frac{\Omega a}{Z'}\ln \Omega - \frac\Omega{Z'}\left(a\ln a\right) - \frac1{Z'}\left( \sum_i^n b_i \ln b_i \right) - S_0\nb\\
= \left(\frac1{Z'}-1\right)S_0 + \ln Z' - \frac{a}{Z'}\Omega\ln\Omega - \frac{(\Omega-1)a}{Z'}\ln a \nb\\
= \left(\frac1{Z'}-1\right)\left( S_0 + \ln a \right) + \ln Z'  - \frac{a}{Z'}\Omega\ln\Omega
\eqn
Maximizing $S_0$ we have,
\bq
S_0 = -\ln a 
\eq
which eliminates the first term. Then maximizing for $a$  we have
\bqn
a_{max} = (\Omega-1)^{-2}\left( 1 - \Omega + \Omega \ln \Omega\right)\\
Z_{max} = \frac{\Omega\ln\Omega}{\Omega-1}\\
\Delta S_{max} = \ln Z_{max} - Z_{max} + 1 \nb\\
= -\ln(\Omega-1) - \frac{\ln\Omega}{\Omega-1} + \ln\ln\Omega + 1
\eqn
This entropy difference can be exploited to do work. An example of this is a post-selected entangled Szilard engine.
In the classic Szilard engine a particle is trapped in a box that is then divided into two equal chambers. 
An observer then measures which chamber the particle is in, and moves a piston that collapses the opposite chamber. 
When the partition dividing the chambers is removed, the pressure of the gas particle pushes the piston out, doing work. 
The process is then repeated, thus extracting useful work from a single heat bath. 
The hitch in this scenario is that it is not a complete cycle until the bit of information gained by measurement is erased, increasing entropy. 
The entropy cost for operating the engine balances out the work done, preserving the second law of thermodynamics. 
This erasure cost is given by Landauer's principle. 
We can attempt to circumvent this by changing the relative size of the chambers, but on average the cost will at least cancel and often outweigh any work gained.
However, if the observation of the particle is tied to a post-selected variable, then out ensemble can essentially cherry pick out those cases in which we come out ahead. 
In this thought experiment the net work done is
\bq
W = -P_{left}\ln x - P_{right}\ln ( 1-x ) - \ln 2
\eq
The partition is located a distance $ 0 < x < 1$ from the left side of the box. The constant term is from the erasure of the single bit of observation done as an intermediate step in entanglement.
The post-selected probabilities are
\bq
P_{left} = 1 - P_{right} = \frac{\Omega x}{ (\Omega -1)x + 1 } 
\eq
The maximum of $W$ requires solving a transcendental equation, so is best done numerically. An approximation shows that the leading term is of order $\ln\Omega$. It may seem that this bound could be violated in the following way. Suppose the partition is placed in the center, and then one of the pistons is removed to a large distance. The entropy difference between entangled and unentangled ensembles for this state may grow arbitrarily large. However, work will only be done if the particle is in the expanding chamber. By selecting the particle into the smaller chamber after the expansion, it may seem that entropy is reduced by a large amount, but the selection process merely back-propagates to before the insertion of the partition, rather than force any sort of 'tunneling' type event. Because of this, no additional work is done. It would be more appropriate to say that a logarithmic work bound exists.  Another way to understand this effect is to consider Crook's fluctuation theorem. 
\bq
P_{A\rightarrow B}( W) = P_{A \leftarrow B}( -W ) \exp\left[ \beta (W - \Delta F)\right]
\eq
The ratio of probabilities of transitions between states of different free energies $\Delta F$ is exponential in the work done.
By skewing these probabilities with post-selection we effectively introduce a multiplier $\Omega$ for the skewed probability ratio. This manifests in an effective shift in the free energy of the system of
\bq
W_{TM} = \beta^{-1}\ln \Omega = TdS
\eq
We may refer to this quantity as the negentropy or entropy potential of a time machine. It represents the amount of excess entropy that must be produced by any process that creates an effective time loop.


\section{Measurement and error}

In ordinary quantum mechanics we learn that the wave-function itself is not an observable. 
Probabilities of events are normally only measurable by testing the frequency of the events. 
The reason we have to comb through large data sets to test models that predict very unlikely events or small changes in
the relative probability of more common ones is that we generally have to wait and see to construct a frequentist model.
In post-selected systems the amplitude of some branch of the wave-function is enhanced or suppressed by a particular amount, depending on how
the system is entangled with the periodic channel. By controlling the degree of entanglement we can construct a sort of probability microscope.
This allows us to observe rare processes by renormalizing parts of the wave function that we want to interact more strongly with an observable.
To do this, we employ a 'fuse'. That is a probabilistic algorithm with a known exponentially small success probability.
This fuse is then placed in parallel with the problem we wish to estimate. We can then compare the success rates of the two paths to estimate the success rate of the unknown path.
If the fuse is 'blown' we can place an exponentially small upper bound on the probability of the unknown process.
There is a caveat that unless back-propagation is compensated for in some way, iterative measurements may interfere with each other.\\
Another distinguishing feature of nonlinear modifications to quantum mechanics is the ability to more reliably distinguish non-orthogonal states. 
Any process which allows this typically leads back to the creation of effective CTCs and all the paradoxes of this paper.
 In ordinary quantum mechanics the optimal method for distinguishing between two pure states in 
a mixture using projective measurement is to choose one of the eigenvectors to be one of the two states to be distinguished. Some improvement over this is gained by 
using the technique of positive operator valued measurements. It is a well known result that the three operators
\bqn
\Pi_a = \frac{1 - |b\rr\lr b|}{1 + |\lr a| b \rr |}\\
\Pi_b = \frac{1 - |a\rr\lr a|}{1 + |\lr a| b \rr |}\\
\Pi_n = 1 - \Pi_0 - \Pi_1
\eqn
Allow for the maximum success probability for discerning the two states.
 By acting with a not and two $C_{not}$ gates we can alight the post-selection to project out the 'inconclusive' demension of the expanded POVM space.
 Rather than an inconclusive result probability of
\bq
P_n = | \lr a | b \rr |
\eq
we have
\bq
\bar{P}_n = \frac{P_n}{P_n + \Omega -\Omega P_n}
\eq
Because the probability of a false identification is zero, it is unchanged by the renormalization. The relative ratio of detections is unaffected, only the ambiguous result decreases. 
The mixture of the discerned bit is not otherwise affected, remaining a combination of the three orthogonal outputs. The lack of further back-propagation is due to the symmetry of the measurement process between the $|a\rr$ and $|b\rr$ states.
In a more general case, states in the mixture are renormalized away from the mean. If a third state is added to the mixture it will be suppressed in the ensemble by how close it is to the inconclusive projection. 
For states that form an angle $\theta$ with the renormalized mean state of $|a\rr+|b\rr$,
\bq
P_n(\theta) =\frac{4P_n\cos^2\theta}{2 + \lr a | b \rr + \lr b | a \rr }
\eq
and the relative weight in the ensemble will be skewed by a factor
\bq
w(\theta) = 1-P_n(\theta) + \bar{P}_n(\theta) 
\eq
In the limit as we attempt to discern nearer and nearer states, we re-create a grandfather paradox centered on the mean state, and amplifying small fluctuations away from it.\\
One of the more curious things in measurement theory is the non-uniqueness of the composition of density matrices. 
The density matrix normally contains all the information needed for the prediction of measurement of a system.
 In standard decoherence theory it also represents a degree of ignorance of the system's entanglement with the environment. 
The complementary information resides in the hidden degrees of freedom and the mutual phase of the state components that are traced over to give the density matrix.
For post-selected quantum mechanics density matrices maintain a similar ambiguity as in regular quantum theory. 
Each weight and projection are the same if the unprojected density matrix is the same, and the post-selected matrix is formed from these quantities.  
Unlike the increased state discrepancy, there is no outcome to select against, since the value of no channel can depend on the difference between two equivalent density matrices.\\
The treatment of weak measurements is thoroughly covered elsewhere, and being designed for post-selected systems, needs no modification\cite{ahr1}.
 In the presence of noise, we add to the ensemble the weak result of projecting against each orthogonal direction for the Bell state or classical channel, 
and combine the sub-ensembles with a relative weight of $\Omega$ multiplying the desired projection's noiseless result. 
Suppose a weak measurement is given by an operator $A$ inserted into the circuit between two unitary evolutions. Normally we would have,
\bq
\lr A \rr_{weak} =  \lr \psi_{in} | U_2  A U_1 |\psi_{out}\rr
\eq
for exact post-selection. The expectation value with noise is simply the weighted average of multiple weak measurements.
\bq
\lr A \rr_\Omega = \frac1{Z}\left( \lr \phi_\perp | U_2 A U_1 | \phi \rr + \Omega\lr \phi | U_2 A U_1 | \phi \rr \right)
\eq

The suppression of ambiguous measurement outcomes can also be used to improve error correction. 
An example of this is a two qubit scheme, in which we periodically measure the parity of two entangled qubits, and select against odd states.
Two Bell states, one being the periodic channel, and the other the communication channel where we have a small probability of $\epsilon$ of flipping a bit.
We take the noisy product state to be
\bq
|\psi_{out}\rr = |\phi_{TM}\rr \otimes \left( (1-\epsilon) \left( \alpha|00\rr + \beta|11\rr \right) + \sqrt{\epsilon(1-\epsilon)} \left( \alpha + \beta\right)\left( |01\rr + |10\rr\right) + \epsilon\left( \beta |00\rr + \alpha |11\rr\right) \right)
\eq
Then we use two cnot gates one from each part of the second Bell channel.
\bq
|\psi_{in}\rr = U_{cnot32}U_{cnot42}|\psi_{out}\rr
\eq
Then the projected states are
\bqn
|\bar\psi_B \rr =   \left(\alpha(1-\epsilon) + \beta\epsilon\right)|00\rr +  \left(\beta(1-\epsilon) + \alpha\epsilon \right)|00\rr \\
|\bar\psi_N \rr = (1-\epsilon)\epsilon(\alpha + \beta\left( |01\rr +  |10\rr \right) 
\eqn
The partition function will be
\bq
Z = \Omega\lr \bar\psi_B | \bar\psi_B\rr + \lr \bar\psi_N | \bar\psi_N \rr
\eq
\begin{figure}[h!]
  \caption{A scambler $U_t$ and selection based error correction.}
  \centering
    \includegraphics[width=0.4\textwidth]{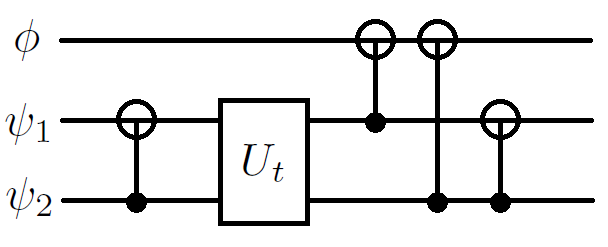}
\end{figure}

In the limit of $\Omega\epsilon>>1$ it appears that we can correct $n$ errors with only $n+1$ qubits, up to the limit of fidelity
\bq
f_n =\frac{(1-\epsilon)^{n+1}}{(1-\epsilon)^{n+1} + \epsilon^n}
\eq
If $\Omega$ is smaller, the ideal correction scheme depends upon it's value relative to the number of qubits and the individual error rate $\epsilon$. 


\section{Tourist trap}
In a famous criticism of the science fiction of time travel, a few authors such as Hawking have remarked about the distinct lack of tourists from the future. The idea is that eventually someone will want to visit a particular moment, so we should immediately see them. The logic is similar to a combination of the Fermi and Obler paradoxes. Resolutions follow several avenues. First the asymptotic behavior of this 'time traveler expectation value' may easily yield a finite, even small result if it decays quickly enough. The work required to create a channel is extensive in the number of qubits, and with a low error rate it may be extremely large. At constant temperature the normalization factor for a loop in a thermal bath typically falls of exponentially. \\
To set up a model circuit for this problem will require several channels. In some capacity we are looking at a slightly more complex version of the unproven proof scenario, but with another twist, we want to model a conditionally post-selected channel. One channel we will designate $|p\rr$, a qubit corresponding to whether the periodic channels are 'on'. Next we will use two sets of $n$-qubit control channels initialized to the $|0\rr^n$ state. One qubit from each set of control channels will act on a single qubit from the set of periodic channels, creating an $n$-qubit version of the third party paradox circuit, with all null inputs.To send a message, a set of $n$ parallel Toffoli gates will conditionally copy a message onto one of the control channel sets if the power qubit is on. To receive the message a set of $n$ Bell pairs is used. One member of each pair is conditionally copied to the other set of control channels with a similar set of Toffoli gates, again taking the power qubit as one of the two controls in each gate. The Bell partners will, assuming post-selection is successful and the power qubit is on throughout, effectively relay the message.  If we wish, we can use the Bell partners as the sending channel, or act on it with gates in between. If we want to simulate the idea of a time traveler bringing us the plans for the time machine itself, we can condition the value of the power qubit on the message received, as the receiving channel is measured before either set of Toffoli gates acts. Lastly we can post-select to give the power qubit an independent skew factor, representing an arbitrary degree of waste entropy production, or other back-reaction effects. The initial product state is composed of the periodic pairs, the receiving pairs, the power qubit, and the sending qubits.
\bq
|\psi_{out}\rr = |\phi_{Bell}\rr^n \otimes |p\rr \otimes | m \rr \otimes | \psi_{Bell} \rr^n
\eq
We can relate the message to be sent to the received message by a unitary evolution of received message and the environment.
\bq
|m\rr = U_t ( |\psi_{bell}\rr^n \otimes |e\rr )
\eq
The second members of the receiving Bell pairs can be traced over to simulate the decoherence of the received classical message due to the interaction with the rest of the world.
The normalization factor will be a sum over the projections against of each of the $2^{2n}$ orthogonal basis states for the $n$-qubit periodic channel, weighted by the number of errors. For a number of errors $n-q$ we have
\bq
\bar\psi_q = \lr \phi^q \otimes \phi_{\perp}^{n-q} | U_t | \phi_{bell}^n\rr 
\eq
and therefore partition function of,
\bq
Z = \sum_{\bar\psi} \epsilon^n\Omega^q \lr \bar\psi_q|\bar\psi_q\rr
\eq
where $\epsilon$ is the per bit error rate. This is assuming all states are projected. But if only some are projected it becomes more complicated. 
Effectively we are no longer projecting against the rotated Bell states, but against the product of those states and the power qubit. For projections where the power qubit is off the skew factor between the different orthogonal Bell states is unity. For the single qubit conditional periodic channel we have eight possible projections, and so eight possible weights. If we use $\omega_{on/off}$ for relative weight of normal and noisily post-selected subensembles,   
the effective partition function will be
\bq
Z_p = \omega_{on}\left(\left(1-\frac34\lambda\right)N^2 + \frac\lambda{4}\right) + \omega_{off}
\eq
Including members in the ensemble that are not post-selected effectively increases the noise.
\bq
\Omega_p = \frac{\omega_{on}}{\omega_{off}+\lambda/4}\Omega_\lambda\cdot\frac\lambda{4}
\eq
This trade off can be seen in the third part paradox circuit if we add an additional rotation gate to the periodic channel. Both noise and any conditionality on the projection add a representative unbiased subensemble to our post-selected ensemble, decreasing skew effects.\\
Suppose that the power channel is set to some function of the message and environment. This could correspond to the recognition of a message separate from random noise, or some other physical prerequisite for post-selection. Because the normalization factor is always less than one, unless we insert an a priori bias for the power channel to be on, it will be set to on strictly less frequently than it would if there were no post-selection. Since only histories where post-selection occurs can be deselected from an ensemble, an arbitrary ensemble will have a bias against producing such events, if such selection is possible. Due to the dissipative nature of decoherence, any 'time tourist' would be exponentially unlikely to satisfy the periodic condition.  Since mutual phase information would be lost when they interacted with the environment, they act like a large $n$ channel with a random distribution of phase flip gates.  The normalization factor of the state corresponding to their appearance decays exponentially with interaction, down to an equilibrium value exponentially small in $n$. Since a thermal bath has a probability of producing any message that is exponentially small in $n$, the vacuum expectation of 'time tourists' may not be significantly different than it would otherwise be for a thermal bath in normal quantum mechanics. Unless the creation of projections and/or time machines is associated with a large increase in entropy, it will be as rare as the corresponding fluctuation of a thermal ensemble that decreases entropy by an amount equal to the entropy potential of the time machine. In order to realize a low error rate for a channel, the normalization factor for a typical message state must be close to unity, otherwise errors will be amplified, and our 'traveler' scrambled.\\
Returning to our toy model circuit, suppose there is a single string that can be formed by the $n$-qubits that will tell us how to build our post-selection/time machine. Since the third party message is received when we measure one of each Bell pair, we need not have built the rest of the circuit yet! What is the probability of getting this message compared to all the other random strings we might find by measuring a bunch of Bell pairs? It is actually $\emph{less}$ than $2^{-n}$. This is because the possibility that we will attempt to send a different message, accidentally or not, will only $\emph{decrease}$ the weight of the histories where the message is received in the first place. One can offset this by adding a skew factor of that history, but this is equivalent to adding a high a priori probability that the particular message will appear. This effect can be thought of as a soft chronology protection principle, that the expectation value of causality violations should drop off at least exponentially in both time and system size, similar to thermodynamic behavior given by the fluctuation theorem.
Another approach to this paradox is to generalize the renormalization process. When we perform a noisy post-selection we are dictating the ratio of particular sub-ensembles to form a skewed ensemble. In this case the ratios of sub-ensembles are given by the projection against the time machine eigenstates and an additional weight, the effective error rate, which prevents our ensemble size from vanishing in general. When looking at projection over larger numbers of states, there is more freedom to assign differences in sub-ensemble ratios. In the Bell model, for example, we could give a different probability for phase errors relative to bit errors in some preferred basis. In multi-qubit channels, there is no a priori reason that the error rates would be the same for different channels. 
Since conditional projection is equivalent to projection against multiple variables, we may also consider limiting back-propagation with a rule. Given two sets of sub-ensembles, one in which a state projection occurs and another where it does not, we may fix the relative probability of the two within the total ensemble, and allow post-selection to only redistribute the weight of histories within sub ensembles that have the same set of 'projection events'. 
Let us proceed to a more definite example.
Suppose we take three Bell pairs and measure one particle of each pair to obtain a classical 3 bit message.
Suppose we have a method of applying some state projection to these partner particles, but only if the first two bits of the classical message are 0. We can then choose the to project the last qubit however we wish, forming a single qubit periodic channel.
The initial state of the three pairs is assumed to be
\bq
|\psi_0\rr = \frac1{\sqrt{8}}\left( |00\rr + |11\rr\right)^3
\eq
Two sub ensembles exist after the measurement, one with the first two bits in the zero state, the other its complement.
\bq
|\psi_1\rr = \frac1{\sqrt{8}}|00\rr \otimes ( |0\rr + |1\rr ) + \frac1{\sqrt{8}}( |01\rr + |10\rr + |11\rr ) \otimes ( |0\rr + |1\rr )
\eq
With the 'insulation rule' in effect, we will have the final projected states
\bqn
|\bar\psi_B\rr = \frac12|00\rr \otimes ( \alpha_B|0\rr + \beta_B|1\rr ) + \frac1{\sqrt{8}}( |01\rr + |10\rr + |11\rr ) \otimes ( |0\rr + |1\rr )\\
|\bar\psi_N\rr = \frac12|00\rr \otimes ( \alpha_N|0\rr + \beta_N|1\rr ) + \frac1{\sqrt{8}}( |01\rr + |10\rr + |11\rr ) \otimes ( |0\rr + |1\rr )\\
|\bar\psi_-\rr = \frac12|00\rr \otimes ( \alpha_-|0\rr + \beta_-|1\rr ) + \frac1{\sqrt{8}}( |01\rr + |10\rr + |11\rr ) \otimes ( |0\rr + |1\rr )\\
|\bar\psi_{-N}\rr = \frac12|00\rr \otimes ( \alpha_{-N}|0\rr + \beta_{-N}|1\rr ) + \frac1{\sqrt{8}}( |01\rr + |10\rr + |11\rr ) \otimes ( |0\rr + |1\rr )
\eqn
While in general this 'third qubit may be entangled with other variables, the premise that the weight in the ensemble of the $|00\rr$ conditional projection state, as well as the relative amplitudes in the other branches of the wave-function, remains independent of the projection we choose. The final state is the normal mixture and partition of these four projections, or two in teh case of a classical channel. While the tourists are not suppressed relative to the thermal bath, they do not become more frequent either. While time machine effects may still back-propagate, they at least do not 'side-propagate' in this model. to be observed in branches of the wave function where there are no time machines. 
Suppose we select against the $|000\rr$ state. Two methods of renormalization are available, depending on if we wish the distribution of sub ensembles to skew the value of the 'power' qubits. Without partitioning we have
\bq
|\bar\psi\rr_{bp} = \frac1{\sqrt{7}}|001\rr + \frac1{\sqrt{7}}\left(|01\rr+|10\rr+|11\rr\right) \otimes \left( |0\rr + |1\rr \right)
\eq
and with partitioning the the projected sub-ensemble we have
\bq
|\bar\psi\rr_{sp} = \frac12|001\rr + \frac1{\sqrt{8}}\left(|01\rr+|10\rr+|11\rr\right) \otimes \left( |0\rr + |1\rr \right)
\eq
In the former case, the probability of conditional post-selection depends on the normalization factor. 
This would give an overall bias against 'tourists', $p_{00}=1/7$, compared with the 'normal' value of $1/4$ for the 'insulated' model, or for no post-selection.\\

\section{Computation}
The new measurement effects allow for some powerful analog computations, but the pack-propagation phenomenon allows for the collapse of the verify-calculate divide in the computational hierarchy directly.
It has been demonstrated independently by several authors that post-selected computation collapses $NP$ to $P$, but it is further the case that this holds for each power individually. For example, verification in quadratic time leads to solutions in quadratic time, lists may be sorted in linear time, and searches conducted instantly.
Another point is that it holds $\emph{for every oracle}$ we might choose to introduce. 
A post-selection assisted computer can physically reverse a calculation process by aligning its outcome with the selection condition. 
As a result it can invert any oracle function, including many very sparse functions that are normally used to divide classes.
A blind password attack can normally only be done in exponential time for completely random keys. Each attempt is modeled as an oracle query, returning yes if the password is correct, and no otherwise.
While not normally considered an interesting problem in complexity theory, it is one of great practical importance.
With post-selection the inverse can be uncovered in effectively linear time. The only reason that the time would not be constant time is due to the amplification of errors.  
An important result by \cite{scott} shows that post-selection efficiently solvesl $PP$ problems in polynomial time almost by definition. $PP$ can be expressed as problems for which a solution can possibly by verified in polynomial time using the best case of a probabilistic algorithm with access to coin flips. There are other definitions of $PP$, but this one makes the connection with post-selection the most clear. The method is we take an ordinary probabilistic algorithm and select along the case that it succeeds. One could call such a method a 'squeezed' probabilistic algorithm, since the normal distribution of the coin flips that the algorithm uses is squeezed onto those combinations of flips that result in an accept. It would be tempting to treat the  system as an oracle for $PP$, but multiple queries can present a problem.
The primary difficulty in extending the individual results is that  treating the squeezed probabilistic algorithms as oracles does not take into account interaction between projections.  In simulating an interactive proof we may use a post-selected probabilistic algorithm to generate the proof to be reviewed, and attempt to use another such algorithm to find errors in the proof,
but adversarial systems will suffer from back-propagation that will cause them to mutually amplify each other's errors, possibly  making the whole program unreliable. 
In order to help insulate against back-propagation, we can employ amnesia type circuits to 'initialize' qubits into a Hadamard state or a classically random state if the channel is classical. Some back-propagation will still occur, but it is due to any singular points that the amnesia circuit has, rather than the action of later gates. In the classical limit, the unproven proof circuit can be used for this purpose, as a 'true random' source. It would be more effective since it has no singular points. \\
One consequence of the ability to efficiently measure small probabilities is that post-selection can efficiently solve some $\#P$ problems. If the number of solutions to a given NP problem is polynomial, we can find all of them in polynomial time, by using a fuse circuit whose success chance is sufficiently small, and iterating through a probabilistic search that tries random strings that have not been tried before until the fuse blows. However these results depend on access to such a low probability variable, with acts as an oracle of sorts. If the fuse is composed of coin flips, it should require polynomial time at least as large as the number of solutions, typically one degree higher.
Another example of this is searching. Given a long unsorted list, we can construct a fuse with a probability of less than $1/n$ of activating, using a logarithmic number of operations. A random guess circuit can move us to a particular random element in a similar number of moves. This suggests that we may find a single element in a list, or verify its absence in log time. These result may seem impressive, but remember, exact post-selection allows unlimited entropy reduction.
Let us explore the search problem with noisy post-selection.
Such hyper searching can only be done reliably if the noise ratio is small enough, that is 
\bq
\Omega > n
\eq
Given a probabilistic algorithm, we can use post-selection to improve its success rate, but only up to a certain factor.
\bq
\bar{P}_s = \frac{\Omega P_s}{\Omega P_s + 1 - P_s }
\eq
Beyond that, we must simply execute the algorithm multiple times. For any fixed finite $\Omega$ we can still only solve problems in BQP in polynomial time.
However, using multiple independent periodic channels we can reduce the probability further. For two noisy selection circuits we have a sort of composition relation.
\bqn
\Omega_{12} \approx \Omega_1 \cdot \Omega_2 \\
\Omega_n \approx \Omega_0^n
\eqn
If a noisy post-selector can perform state projections at a rate linear in time, this suggests that the error rate of a probabilistic algorithm entangled with a series of projections is
\bq
\epsilon_\Omega \approx \frac12(e^{\alpha t}P_0/(1-P_0) + 1)^{-1}
\eq
If we compare this to the Chernoff bound error rate where each test requires a time $\gamma$,
\bq
\epsilon_{chrn} \approx \exp\left[ -2\left(p-\frac12\right)^2 \gamma t\right] 
\eq
For our probabilistic search, the item is found in a single attempt with probabilities
\bqn
P_0 = 1/n\\
p = \frac12 + \frac1{2n}
\eqn
for the two methods respectively. Then the two error rates are
\bqn
\epsilon_\Omega \approx \frac12(1-1/n) e^{-\alpha t}\\
\epsilon_\gamma \approx e^{-n^2\gamma t /2}
\eqn
Repeated projection acts like a heat engine, reducing the entropy of a random list element from $\ln n$ to close to zero, but increasing the entropy of the environment. 
If the time is polynomial in $n$, then the skew factor from a constant rate of projection will be exponential in that polynomial.
\bq
\Omega_{poly} \approx e^{p(n)}
\eq
This also happens to be the minimum nonzero probability bound for PP, or a fuse circuit using a polynomial number of gates. 
Classical probabilistic algorithms should have an entropy bound since classical coin flipping is not reversible. 
State projection is not a reversible process, and should introduce similar bounds for quantum computers that utilize it.
We can characterize probabilistic algorithms not only by time and space limitations but also by entropy production. 
In many ways post-selected computing acts like adiabatic computing given an exponentially long cooling time.


\section{Remarks}

The variety of strange behaviors in quantum theory deformed by projections is at least as large as for ordinary quantum mechanics. If physics truly makes progress through paradoxes, then this is fertile ground. With projection the theory space available to a QFT becomes much larger, but also subject to more experimental constraints. The modeling of collapse toward a final state as a postselected computations has drawn considerable interest\cite{presk}. The creation of secondary channels and other back-propagation effects indicate that the exotic behaviors are not necessarily confined to the black hole interior or high curvature regions in the case of the final state model of black holes. However, the dilution of the skew as one moves away from the projection due to increasing entanglement provides some shielding effect. The diminishing returns of noisy postselection for a heat engine as one attempts to select states of smaller and smaller amplitude would indicate a partial information loss behavior as a compromise that limits the severity of exotic effects. The tourist suppression effect in the partial loss model could provide a Boltzmann like renormalizing factor for microscopic black hole fluctuations. This would also lead to loss of coherence in the UV, affecting the short distance behavior of gravity, and perhaps allowing a renormalization and indicating a minimum temperature for quantum gravity.

\end{document}